\documentclass[twocolumn]{aastex63}
\usepackage{amsmath}
\usepackage{multirow}
\usepackage{breakurl}
\usepackage{hyperref}
\usepackage{textcomp}
\usepackage{cleveref}[2012/02/15]
\usepackage{fancyhdr}
\usepackage{graphicx}
\usepackage{amssymb}
\usepackage{longtable}
\usepackage{float}
\usepackage{dcolumn}

\usepackage[toc,page]{appendix}
\usepackage[caption=false]{subfig}
\usepackage{color, colortbl}
\definecolor{Purple}{rgb}{.9,.9,1}

\crefformat{footnote}{#2\footnotemark[#1]#3}

\newcommand{\kms}{km~s$^{-1}$}
\newcommand{\s}{$\sim$}
\newcommand{\n}{$-$}
\newcommand{\HI}{\ion{H}{1}}
\newcommand{\OI}{\ion{O}{1}}
\newcommand{\OVI}{\ion{O}{6}}
\newcommand{\CII}{\ion{C}{2}}
\newcommand{\AlII}{\ion{Al}{2}}

\newcommand{\SiII}{\ion{Si}{2}}
\newcommand{\SiIII}{\ion{Si}{3}}

\newcommand{\CIV}{\ion{C}{4}}
\newcommand{\CIII}{\ion{C}{3}}
\newcommand{\SiIV}{\ion{Si}{4}}

\newcommand{\tm}{\tablenotemark} 
\newcommand{\tn}{\tablenotetext}

\defcitealias{Fox_2015}{F15}
\defcitealias{bordoloi_2017b}{B17}
\defcitealias{Savage_2017}{S17}
\defcitealias{Karim_2018}{K18}

\shortauthors{Ashley, T. et al.}

\begin{document}

\title{Mapping Outflowing Gas in the Fermi Bubbles:\\ a UV Absorption Survey of the Galactic Nuclear Wind\footnote{Based on observations made with the NASA/ESA Hubble Space Telescope, obtained at the Space Telescope Science Institute, which is operated by the Association of Universities for Research in Astronomy, Inc., under NASA contract NAS5-26555. These observations are associated with program 15339.}}

\author[0000-0002-6541-869X]{Trisha Ashley}
\affiliation{Space Telescope Science Institute, 3700 San Martin Drive, Baltimore, MD 21218}
\email{tashley@stsci.edu}

\author[0000-0003-0724-4115]{Andrew J. Fox}
\affiliation{AURA for ESA, Space Telescope Science Institute, 3700 San Martin Drive, Baltimore, MD 21218}
\email{afox@stsci.edu}

\author[0000-0003-1892-4423]{Edward B. Jenkins}
\affiliation{Princeton University Observatory, Princeton, NJ 08544, USA}

\author[0000-0002-0507-7096]{Bart P. Wakker}
\affiliation{Department of Astronomy, University of Wisconsin-Madison, 475 North Charter Street, Madison, WI 53706, USA}

\author[0000-0002-3120-7173]{Rongmon Bordoloi}
\affiliation{Department of Physics, North Carolina State University, 421 Riddick Hall, Raleigh, NC 27695-8202}

\author[0000-0002-6050-2008]{Felix J. Lockman}
\affiliation{Green Bank Observatory, P.O. Box 2, Rt. 28/92, Green Bank, WV 24944, USA}

\author{Blair D. Savage}
\affiliation{ Department of Astronomy, University of Wisconsin-Madison, 475 North Charter Street, Madison, WI 53706, USA}

\author[0000-0002-5652-8870]{Tanveer Karim}
\affiliation{Center for Astrophysics, Harvard and Smithsonian, 60 Garden Street, Cambridge, MA 02138, USA}

\begin{abstract}
Using new ultraviolet (UV) spectra of five background quasars from the Cosmic Origins Spectrograph on the 
\emph{Hubble Space Telescope}, we analyze the low-latitude ($|b|$=20--30\degr) regions of the Fermi Bubbles, the giant gamma-ray emitting lobes at the Galactic Center. We combine these data with previous UV and atomic hydrogen (\HI) datasets to build a comprehensive picture of the kinematics and metal column densities of the cool outflowing clouds entrained in the Fermi Bubbles. We find that the number of UV absorption components per sightline decreases as a function of increasing latitude, suggesting that the outflowing clouds become less common with increasing latitude. The Fermi Bubble \HI\ clouds are accelerated up to  $b\sim7\degr$, whereas when we model the UV Fermi Bubbles clouds’ deprojected flow velocities, we find that they are flat or even accelerating with distance from the Galactic center. This trend, which holds in both the northern and southern hemispheres, indicates that the nuclear outflow accelerates clouds throughout the Fermi Bubbles or has an acceleration phase followed by a coasting phase. Finally, we note the existence of several blueshifted high-velocity clouds at latitudes exceeding \s30\degr, whose velocities cannot be explained by gas clouds confined to the inside of the gamma-ray defined Fermi Bubbles. These anomalous velocity clouds are likely in front of the Fermi Bubbles and could be remnants from past nuclear outflows. Overall, these observations form a valuable set of empirical data on the properties of cool gas in nuclear winds from star-forming galaxies.\\
\end{abstract}

\section{Introduction}
The Fermi Bubbles are giant bipolar structures that originate from the center of the Milky Way. They each extend \s12 kpc or \s55\degr\ above and below the Milky Way's disk in gamma ray emission \citep{Ackermann_2014} and have counterparts in multiple wavelengths including X-ray emission, polarized radio emission, microwave emission, and UV absorption \citep{Bland_Hawthorn_2003, Su_2010,  Dobler_2010, Ackermann_2014, Dobler_2008, Carretti_2013, Fox_2015, bordoloi_2017b, Savage_2017, Karim_2018}. The Fermi Bubbles are an example of a nuclear wind within our own galaxy, consequently we can study them in detail and use them analogs for feedback arising from similar mechanisms found in extragalactic sources \citep[for a review of extragalactic and Galactic cool gas flows see][]{Veilleux_2020}.

The origin of the Fermi Bubbles has been debated since their discovery, with two alternative energy sources invoked: the central supermassive black hole (Sgr A*) and star formation near the center of the Galaxy. Both mechanisms could supply the required energy to form the bubbles, but on very different timescales. In the Sgr A* origin scenario, the Bubbles formed in the last few Myr by either an accretion-disk wind or jet activity \citep{Zubovas_2011, Guo_2012, Yang_2012, Bland_Hawthorn_2013, Mou_2014, Bland_Hawthorn_2019}. In the star formation-driven scenario, the Bubbles were formed by powerful star formation near the center of the Galaxy over the past \s100 Myr \citep{Yusef_Zadeh_2009, Lacki_2014, Crocker_2015}. 

While it is possible that both Sgr A* and star formation contributed to the formation of the Bubbles, multiple lines of evidence now point toward Sgr A* as the primary origin because it explains the energetics on physically plausible timescales. Specifically, UV absorption-line and \HI\ kinematic studies of clouds traveling within the Fermi Bubbles indicate a wind age of \s6-9 Myr \citep{Fox_2015,bordoloi_2017b, Di_Teodoro_2018}, the same order of magnitude as Sgr A* formation model timescales, such as the \citet{Guo_2012} jet formation models and \citet{Zubovas_2011} accretion event models with timescales of 1-3 Myr and \s6 Myr, respectively. Furthermore, \citet{Bland_Hawthorn_2013, Bland_Hawthorn_2019} find elevated levels of H$\alpha$ emission and UV ionization in the Magellanic Stream below the Galactic Center, consistent with the ionization cone that would be expected in a Seyfert flare event several Myr ago. Finally, a $\sim$100\,Myr extended period of star formation would have produced a large B-star population at the Galactic Center that is not observed unless an unusually top-heavy initial mass function is invoked. 

UV absorption-line data are essential for measuring the kinematics, metal content, and ionization levels of the cool gas ($T_{\rm gas}\sim10^4-10^5$ K) entrained in the Fermi Bubbles \citep[][]{Fox_2015}. Several studies have measured the UV absorption features of the Fermi Bubbles using sightlines to background quasars and, in two cases, a background star \citep[][hereafter, F15, B17, S17, and K18]{Fox_2015, bordoloi_2017b, Savage_2017, Karim_2018}. The model-dependent results of these UV studies have revealed that the Fermi Bubbles have a cool-gas mass of $\sim2\times10^6$ M$_{\sun}$, a mass outflow rate of $\sim$0.2--0.3 M$_{\sun}$\,yr$^{-1}$, an outflow velocity of \s800--1000 \kms, and a wind age of \s6-9 Myr \citepalias{Fox_2015, bordoloi_2017b}.  

\begin{figure*}[!ht]
    \centering
    \epsscale{1.2}
      \plotone{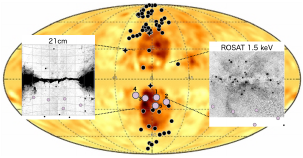}
\caption{The background image is a gamma ray map of the Fermi Bubbles at energies of 3-10 GeV from \citet{Ackermann_2014}. All quasar sightlines through and in close proximity to the Fermi Bubbles are plotted on top of the image. The black circles are quasar sightlines from previous UV studies \citepalias{Fox_2015, bordoloi_2017b, Karim_2018}. The black stars are stellar background sightlines from \citetalias{Savage_2017} and \citetalias{bordoloi_2017b}. The purple circles are five new low-latitude quasars from \citet{Monroe_2016} and whose UV data are presented in this paper. Labels for each of the new pointings are given in Table~\ref{table:observations}. Similar to Figure 1 of \citet{Bland_Hawthorn_2019}, we display the 21 cm map of the Galactic center along the Galactic tangent points from \citet{Lockman_2016} to the left and the 1.5 keV X-ray emission map of the Galactic Center from \citet{Bland_Hawthorn_2003} on the right.}\label{figure:previous_work}
\end{figure*}

 The Fermi Bubbles cover such a large angular extent on the sky ($\pm$55$\degr$ in latitude and $\pm$20$\degr$ in longitude) that there are multiple background sources with sufficient UV continuum for absorption studies using \emph{Hubble Space Telescope (HST)} Cosmic Origins Spectrograph (COS). However, past surveys have been limited to relatively high latitudes ($|b|>30\degr$) because foreground interstellar dust in the disk of the Milky Way extinguishes the light from background quasars (see Figure~\ref{figure:previous_work}). Recently, \citet{Monroe_2016} published a catalog of new UV-bright quasars which includes five new low-latitude sightlines that cover a large range of longitudes through the Fermi Bubbles ($-26\degr \leq b \leq -16\degr$). These new sightlines allow us to probe regions much closer to the launch point of the Bubbles than before in UV absorption and to search for variations in the Fermi Bubbles with Galactic longitude.

In this paper we present HST/COS data of these five new low-latitude quasar sightlines and combine them with previous UV samples \citepalias{Fox_2015, bordoloi_2017b, Savage_2017, Karim_2018} to give a comprehensive view of the UV absorption characteristics of the cool gas in the Fermi Bubbles. In Section~\ref{section:obs_data} we discuss the observations and data reduction and the selection of high velocity clouds (HVCs) associated with Galactic center activity. HVCs are defined as clouds moving at velocities in the Local Standard of Rest (LSR) of $|v_{\rm LSR}|\gtrsim$90 \kms, which is too fast to be co-rotating with the Milky Way disk \citep{Wakker_1997}. Throughout the paper we will refer to the high-velocity clouds detected in the direction of the Fermi Bubbles as Fermi Bubble high-velocity clouds (FB HVCs) to distinguish them from other Galactic HVCs which are not associated with the Fermi Bubbles. We detail the results of the kinematics, ion column densities, and ion ratios of the FB HVCs in Section~\ref{section:results}.  Next, we discuss the results in Section~\ref{section:discussion}. We conclude our paper in Section~\ref{section:conclusions}.

\section{Observations and Methods} \label{section:obs_data}

\subsection{Far-Ultraviolet Data}
The new data analyzed in this paper were taken with \emph{HST}/COS using the G130M and G160M far-ultraviolet (FUV) gratings. The absorption lines covered by this wavelength range include: \ion{O}{1} 1302, \ion{C}{2} 1334, \ion{C}{4} 1548, 1550, \ion{Al}{2} 1670, \ion{Ni}{2} 1317, 1370, 1454 \ion{Si}{2} 1190, 1193, 1260, 1304, 1526, \ion{Si}{3} 1206, \ion{Si}{4} 1393, 1402, \ion{S}{2} 1250, 1253, 1259, \ion{Fe}{2} 1144, 1608, and \ion{N}{5} 1238, 1242.
The data were taken in a total of 22 \emph{HST} orbits and have a signal-to-noise (S/N) of 20 per resolution element.  General information on the UV-bright background sources is given in Table~\ref{table:observations}.

\begin{deluxetable*}{ccccccccc}
\tablecaption{Background quasar Properties and \emph{HST}/COS Observations\label{table:observations}}
\tablehead{\colhead{Sightline\tm{a}} & \colhead{Other} & \colhead{$l$} & \colhead{$b$} & \colhead{$z$}  &  \colhead{FUV\tm{b}}  &  \colhead{Orbits\tm{c}}  &  \colhead{Orbits\tm{c}} & \colhead{Figure}\\[-6pt]
\colhead{} & \colhead{Identifiers}& \colhead{(\degr)} & \colhead{(\degr)} & \colhead{} & \colhead{(mag)} & \colhead{G130M} & \colhead{G160M} & \colhead{Label}}
\startdata
UVQS J185302-415839  & & 354.36  &  $-$18.04  &  0.1842  &  17.32  &  4  &  4 & 1\\
UVQS J185649-544229  & & 341.66  &  $-$22.60  &  0.0570  &  16.39  &  2  &  2 & 2\\
UVQS J191928-295808  & PKS 1916-300 & 8.18    &  $-$18.77  &  0.1668  &  16.26  &  2  &  2 & 3\\
UVQS J192636-182553  & & 20.00   &  $-$15.86  &  0.0786  &  15.67  &  1  &  1 & 4\\
UVQS J193819-432646  & IRAS F19348-4333 & 355.47  &  $-$26.41  &  0.0787  &  16.53  &  2  &  2 & 5\\
\enddata
\tn{a}{Name of the background quasar from \citet{Monroe_2016}.}
\tn{b}{GALEX FUV magnitude (AB system).}
\tn{c}{Number of \emph{HST}/COS orbits UV-bright background sources were observed for under Program ID 15339.}
\tablecomments{The data can be accessed at MAST via the following link:  \url{http://dx.doi.org/10.17909/t9-ndq9-cw90}.}
\end{deluxetable*}

The data were reduced with the \emph{calcos} calibration pipeline twice, once using all photons collected and once using only the photons collected taken during orbital night; this ``night-only" reduction reduces geocoronal airglow emission, which can strongly affect \ion{O}{1} $\lambda$1302. After \emph{calcos} was run, a further set of customized velocity alignment and co-addition steps was applied, following the procedures described in detail in the Appendix of \citet{Wakker_2015}. This procedure aligns the Galactic absorption lines detected with the atomic hydrogen (\HI) 21\,cm emission-line data.

The data were binned by five pixels and each absorption line was identified and measured using a Voigt profile fitting program, VPFIT \citep{Carswell_2014}. Initial guesses for fit parameters (line centroid, column density, linewidth, and number of components) were first made using RDGEN, a collection of routines that is included with VPFIT. VPFIT then uses those initial guesses and finds the best fit parameters for each component based on the chi-squared value of the fit, and will reject components that are not necessary to improve the goodness of fit. The wavelength-dependent COS line-spread functions were used in VPFIT to account for instrumental line broadening. Each fit was visually inspected. 

While our fits to the data include absorption components at all velocities, our analysis is focused on the Fermi Bubble high-velocity clouds. In Fermi Bubble directions, the HVC kinematics \citepalias{Fox_2015, Savage_2017} and the over-abundance of HVCs \citepalias{bordoloi_2017b} compared to non-Fermi-Bubble directions supports the physical association of the HVCs with the Fermi Bubbles. We securely identified FB HVCs in the five new sightlines by only accepting HVCs that are detected in two or more metal lines. For this paper, we decided to use a novel approach of matching velocity centroids in a statistically rigorous manner. In order for two absorption features to be labeled as part of the same FB HVC, we required their velocity centroids to match within 3 times their errors, where the errors are given by VPFIT. We discuss this error analysis in detail and the consequences of it in Section~\ref{vel_matching} and Appendix \ref{appendix:centroid_offset}. 

\begin{figure*}[ht!]
    \centering
     \epsscale{1.15}
     \plotone{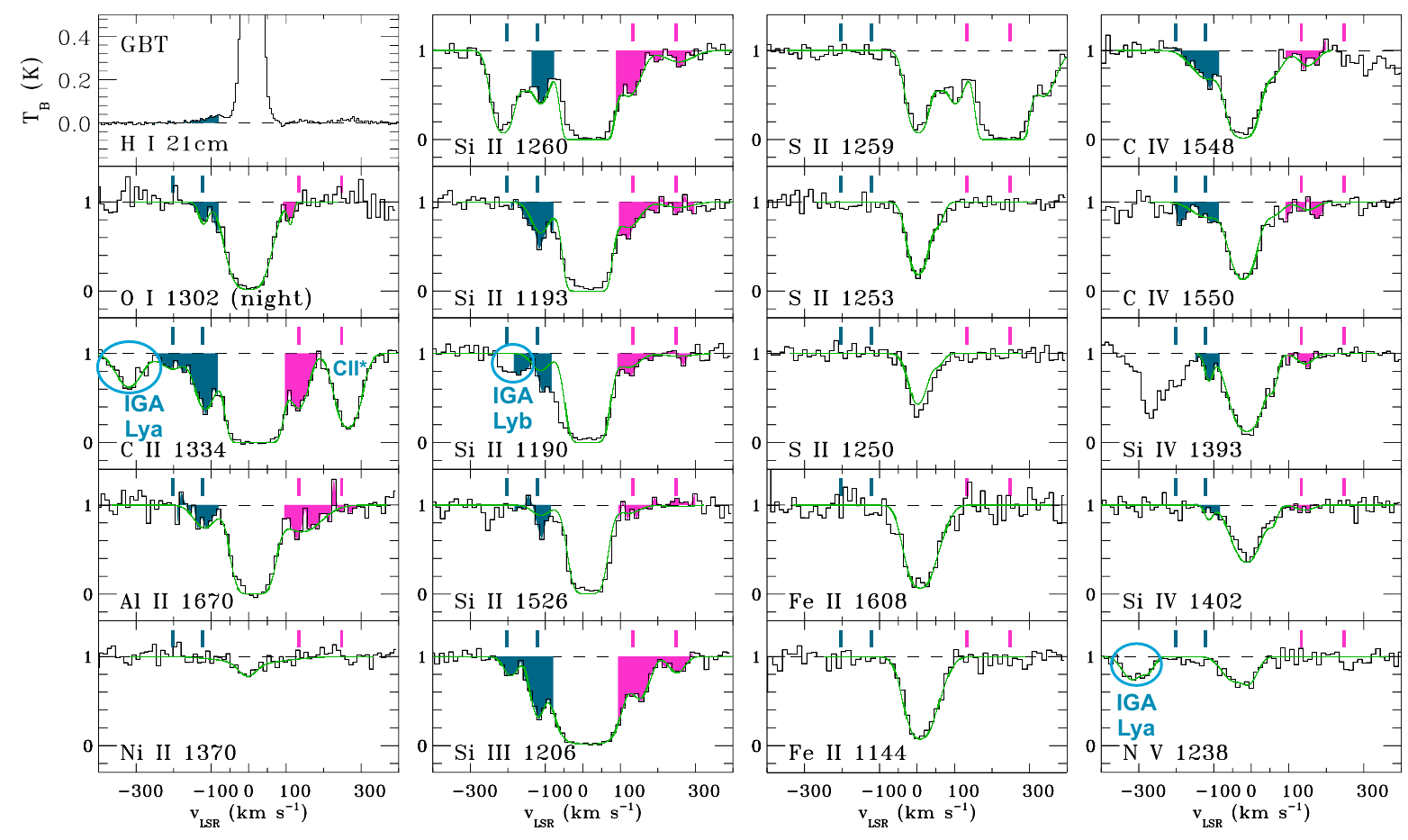}
     \caption{UV absorption-spectra from \emph{HST}/COS of the quasar UVQS J185302-415839. The observations were taken with the G130M and G160M gratings. The 21\,cm \HI\ emission profile in this direction from the GBT is plotted in the top-left panel. 
     The black and green lines represent the data and the fit to the data, respectively. The highlighted pink and blue regions indicate potential FB HVC detections. The median velocity of each FB HVC is indicated by the pink and blue dashes in each panel for reference. Intergalactic absorbers (IGA), Lyman-alpha (Lya) and Lyman-beta (Lyb), and absorbers intrinsic to the background quasar are circled and labeled in each ion plot. Spectra for the other four new sightlines studied in this paper are shown in Figure~\ref{figure:remaining_fits} of the Appendix. \label{figure:fits}}
\end{figure*}

An example of the calibrated data and resulting fits for UVQS J185302-415839 is shown in Figure~\ref{figure:fits}; the fits to the four remaining new sightlines are presented in the Appendix as Figure~\ref{figure:remaining_fits}. Intergalactic absorbers and absorption features inherent to the quasar are identified in each sightline and marked when visible in Figures~\ref{figure:fits}~and~\ref{figure:remaining_fits}. 

\subsection{Atomic Hydrogen 21 cm Data}
For reference, atomic hydrogen (\HI) 21\,cm emission-line data from the Green Bank Observatory (GBT18A-221, PI: Fox) are included in the top-left panel for each sightline in Figure~\ref{figure:fits} when available. Spectra of the 21cm \HI\ line were obtained toward four of the five quasars using the 100 meter diameter Robert C. Byrd Green Bank Telescope (GBT) of the Green Bank Observatory\footnote{The Green Bank Observatory is a facility of the National Science Foundation, operated under a cooperative agreement by Associated Universities, Inc.}. The quasar UVQSJ185649-544229 is below the horizon at Green Bank, so data from the Parkes Galactic All Sky Survey \citep[GASS; ][]{McClure_Griffiths_2009} were used instead. The GASS data have been rebinned by 5 channels.

The angular resolution of the GBT observations is $9\farcm1$.  Spectra were taken by frequency switching, were smoothed, calibrated, and corrected for stray radiation following the technique of \citet{Boothroyd_2011}, and a polynomial was fit to emission-free regions of the spectra to remove residual instrumental baseline.  The polynomial order ranged from 3 to 7. The resulting spectra cover velocities $-748 \leq V_{\rm LSR} \leq +732$ km s$^{-1}$ at a velocity resolution of 1 km s$^{-1}$.  The rms brightness temperature noise in a 1 km s$^{-1}$ channel is 9.5 mK for all sources except for UVQSJ193819-432626, where it is 11.5 mK.  The resulting $5\sigma$ limit on the detectable HI column density, N$_{HI}$, for a 25  km s$^{-1}$ FWHM line is $4.3 \times 10^{17}$ cm$^{-2}$ except for J193819-432624, where it is $5.2 \times 10^{17}$ cm$^{-2}$. The \HI\ spectra displayed in Figure~\ref{figure:fits} have been smoothed to a velocity resolution of \s5.3 \kms. Analysis of the HI spectra will be the topic of a separate paper. 

\subsection{Velocity Centroid Matching}\label{vel_matching}

Typically, absorption features are assumed to be real if they appear in two or more ions with matching velocity centroids and equivalent widths that are consistent with each line's oscillator strength. How closely the velocity centroids of the absorption features have to match in order for them to be considered part of the same cloud is often open to interpretation. In this paper we quantified the proximity of velocity centroids and used these measurements to decide which features are likely living together in the same cloud.  

VPFIT's output includes a velocity centroid and velocity centroid error for each absorption feature fit.  Additionally, there is an intrinsic zeropoint error of $\approx$7.5~\kms\ in each COS FUV grating \citep{Plesha_2019}, which we add in quadrature to the velocity centroid error.  We use these errors to match absorption features to each other within the same sightline.  

First, we separate the absorption features into large groups based on their velocity centroids.  Generally, the range of velocity centroids in a grouping attempts to capture absorption features that visually appeared to align in plots. However, we note that if a single FB HVC has two components separated by a large velocity, then they would be treated as separate clouds in these initial groupings. We assigned groups to the remaining absorption features based on their closest velocity centroid group. These initial groupings are shown in the Appendix Table~\ref{table:detected_lines} as separated by horizontal lines.  

The absorption features within each initial grouping are then matched based on their velocity errors.  The velocity centroid offsets and errors between two ions are calculated using the following equations: 

\begin{equation}\label{eq:change_vel}
\Delta v=v_{2}-v_{1}
\end{equation}

\begin{equation}\label{eq:vel_err}
\sigma_v=\sqrt{(\epsilon_{v,2})^{2}+(\epsilon_{v,1})^{2}}
\end{equation}
where $v_{2}$ and $v_{1}$ are the velocity centroids of two different absorption features within the same initial group and $\epsilon_{v,2}$ and $\epsilon_{v,1}$ are their respective errors. The COS zeropoint errors are only added to $\epsilon_{v}$ (in quadrature) when the absorption features appeared on different gratings. 

Next, the absorption features are matched to each other by comparing the results of equations~\ref{eq:change_vel} and \ref{eq:vel_err}: 

\begin{equation}\label{eq:no_match}
\mathrm{If}\  \Delta v\ \begin{cases} > N\sigma_{v}\mathrm{,}\ \mathrm{then\ no\ match}\nonumber\\[5pt]
< N\sigma_{v}\mathrm{,}\ \mathrm{then\ match}\nonumber
\end{cases}
\end{equation}

where we selected $N=3$ to yield 3$\sigma$ matches. We tested larger values of $N$, however, we found that values greater than 3 resulted in separate absorption features from the same ion in the same pointing beginning to match. We therefore used $N=3$ as a conservative value. Choosing a value of $N=3$ also does not exclude a significant number high-low ion pairs within an initial group that have velocity offsets, as is discussed in detail in Appendix~\ref{appendix:centroid_offset}. Equations~\ref{eq:change_vel} and \ref{eq:vel_err} were used to find matches between every absorption feature in a group, resulting in a grid of matched and unmatched absorption features. 

Finally, when all of the matches between individual ion pairs are made, all of the matched components within a visually aligned initial group are compared to decide which ones are part of the same cloud. This step is important for two cases: (1) absorption features with large velocity errors and (2) ions that have centroids matching some, but not all of the other ion centroids in a group. We do this by starting with the ion that has the lowest velocity centroid error and at least one matched ion. Then, we find all of the ions that are matched to it and also matched to each other (starting with the lowest velocity centroid errors and moving towards increasingly larger centroid errors). These ions are then grouped together as part of the same cloud.  Ions that match some, but not all of the ions in a cloud are marked as potentially being part of that cloud but we note that the result is inconclusive; these absorption features are not used in the calculations or graphs throughout the paper unless specifically stated otherwise. Column 10 of the Appendix Table~\ref{table:detected_lines} indicates to which cloud each ion belongs. The matched absorption features are used to make all figures and calculations throughout this paper for all UV samples. 

\section{Results}\label{section:results}

Of the five new Fermi Bubble sightlines, three have FB HVCs detected in UV absorption. The directions without a FB HVC detection, UVQS J192636-182553 and UVQS J185649-544229, lie just outside and close to the edge of the gamma-ray-defined southern Fermi Bubble, respectively. In Table~\ref{table:HVCs} we present basic information on the FB HVCs detected, including their velocity centroids and list of ions detected. In the Appendix, we provide a full table of fitted line parameters for Fermi Bubble high-velocity absorbers. 

\begin{deluxetable*}{lccclcl}
\tablecaption{FB HVC Detections\label{table:HVCs}}
\tablehead{\colhead{Sightline} & \colhead{$l$} & \colhead{$b$} & \colhead{Allowed Velocities\tm{a}} & \colhead{$v_{\rm LSR}$\tm{b}} & \colhead{error} & \colhead{Ions Detected\tm{c}} \\[-6pt]
\colhead{} & \colhead{($\degr$)} & \colhead{($\degr$)} & \colhead{(\kms)} & \colhead{(\kms)} & \colhead{(\kms)} & \colhead{}}
\startdata
UVQS J185302-415839  &  354.36  &  $-$18.04  & $-$49 to 0 &  $-$121.9  &  9.2  &  OI, C II, Al II, Si II, Si III, C IV, Si IV\\
  &    &    &  & $-$202.4  &  9.9  &  C II, Si III\\
  &    &    &   & \ \ 133.4   &  10.9  &  C II, Al II, Si II, C IV, Si IV\\
  &    &    &   & \ \ 248.0   &  9.4  & Si II, Si III\\
UVQS J191928-295808   &  8.18   &  $-$18.77   & 0 to 63 & \ \ 234.4   &  8.9   &  Si III, C IV\\
UVQS J193819-432646   &  355.47   &  $-$26.41  & $-$13 to 0 &  $-$104.4   &  8.5   &   CII, Si III, Si IV\\
\enddata
\tn{a}{Range of velocities allowed for gas co-rotating with the disk, calculated following the methods outlined in \citet{Wakker_1991}.}
\tn{b}{The average velocity centroid of the matched ions in the last column.}
\tn{c}{A list of matched ions that were found to be part of the same cloud via the process described in Section~\ref{vel_matching}.}
\end{deluxetable*}

\subsection{Incidence of Absorbers}
\begin{figure*}[ht!]
    \centering
    \epsscale{0.92}
      \plottwo{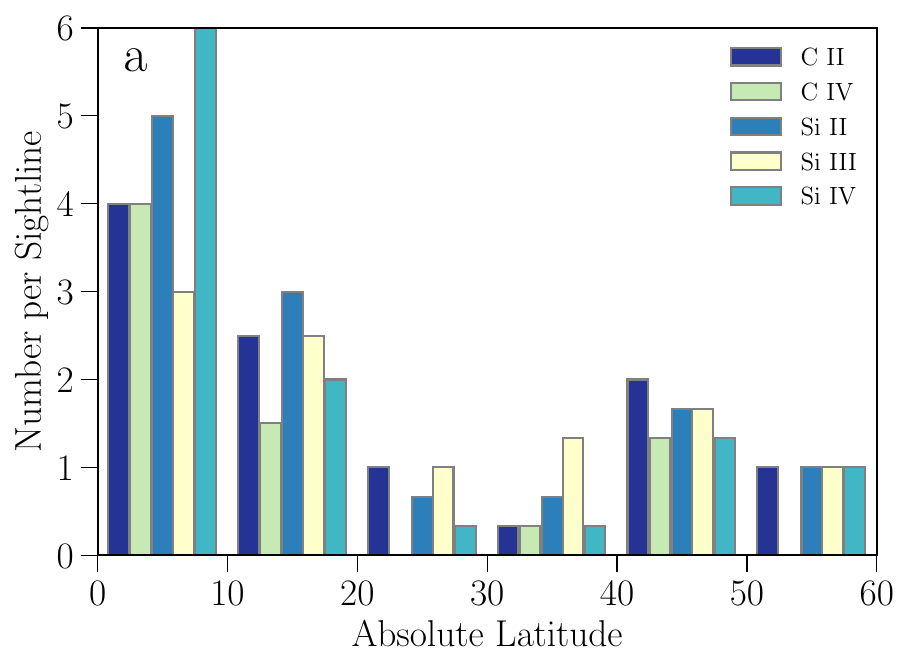}{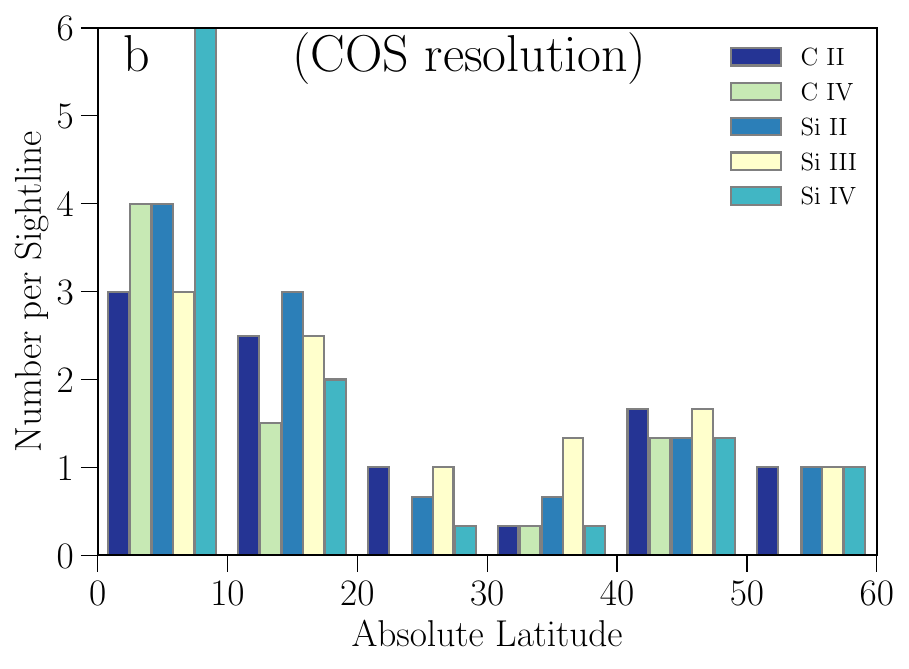}
          \epsscale{0.47}
      \plotone{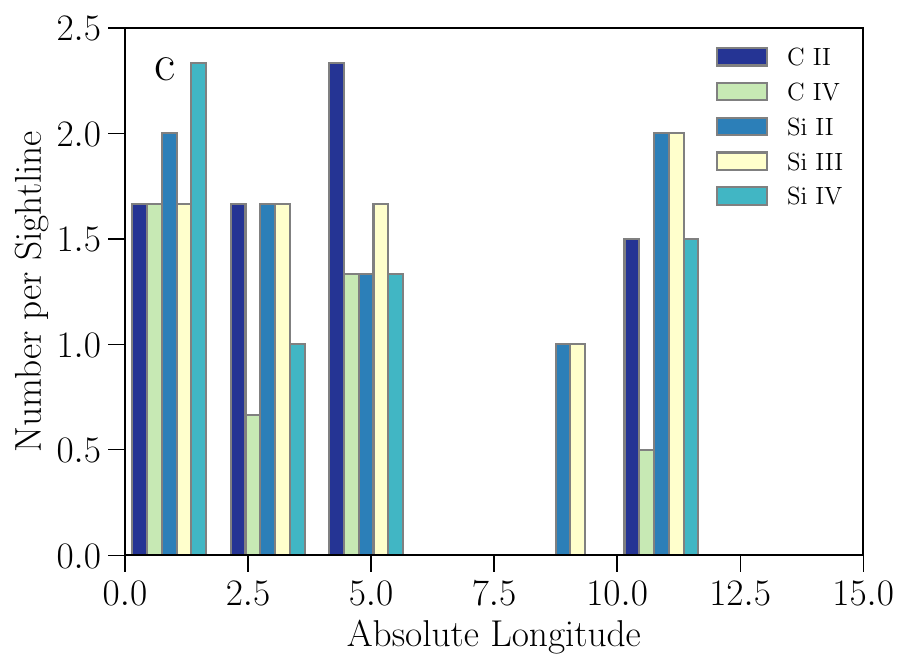}
      \caption{Plots of the number of FB HVC components per sightline as a function of position for \CII, \CIV, \SiII, \SiIII, and \SiIV\ in the full FB sample. (a) All components are plotted vs. absolute latitude. (b) Same as (a) but with STIS detected narrow components that would appear as blended with COS resolution, reduced to one component. (c) All components are plotted vs. absolute longitude.   \label{figure:number_density}}
\end{figure*}

In Figure~\ref{figure:number_density} we plot the number of clouds per sightline detected in various ions against Galactic latitude and longitude to distinguish how the cool gas clouds are distributed in the bubbles. The bin sizes 10\degr\ in latitude and 2\degr\ in longitude. We include our new observations and the samples of \citetalias{Fox_2015}, \citetalias{bordoloi_2017b}, \citetalias{Savage_2017}, and \citetalias{Karim_2018}, forming what we refer to as the ``full Fermi Bubble (FB) sample".

Each ion shows the same general trend with latitude in Figure~\ref{figure:number_density}a: the number of clouds per sightline decreases rapidly beyond the $|b|$ 0-10$\degr$ bin. Between $|b|$ 10$\degr$ and 40$\degr$, there is a decrease in the number of absorbers per sightline. A linear fit to the data between $|b|$ 10$\degr$ and 40$\degr$ results in a slope of \n0.08$\pm$0.02. To test the quality of a monotonic fit to the data between latitudes of 10$\degr$ and 40$\degr$, we also performed a Spearman's Rank Correlation test. The results of this test include the Spearman's correlation coefficient ($\rho_{\text{S}}$) and a two-sided significance of the coefficient's deviation from zero ($P$-value). The correlation coefficient has values from 0 to 1 and a high value of indicates that the data have a strong monotonic relationship. The associated $P$-value indicates how likely it is that any monotonic relationship is due to chance. The data between $|b|$ 10$\degr$ and 40$\degr$ have a $\rho_{\text{S}}$ of \n0.727 and $P$-value of 0.002 indicating that the relationship is strongly monotonic. 

From the $|b|$ 30-40$\degr$ bin to the 40-50$\degr$ bin, there appears to be a small jump in the number of absorbers per sightline. This latitude corresponds to the upper boundary of the Fermi Bubbles, so the increasing incidence at that latitude may be due to the sightlines intersecting interface layers at the Bubble boundaries.

The absorbers in the $|b|$ 0-10$\degr$ bin are all detections from a single STIS/HST sightline, LS 4825, and the $|b|$ 40-50$\degr$ bin contains the STIS/HST sightline M5-ZNG1\footnote{PKS 2005-489 also has STIS and FUSE data, however, the FB HVC from these data at -107 \kms\ was not detected in the silicon or carbon ions in Figure~\ref{figure:number_density} \citep{Keeney_2006}.} \citep{Zech_2008, Savage_2017}. These STIS data have a resolution of 6.6-10 \kms\ while COS data from the rest of the full Fermi Bubble sample have a resolution of 18-20 \kms. To determine if the high resolution data from STIS are resulting in a larger number of absorbers per sightline, we have identified narrow STIS components from LS 4825 and M5-ZNG1 that would not have been resolved as separate features by COS. These components typically have small $b$-values ($\le$10 \kms) and overlap with adjacent absorbers (absorbers within $\le$15 \kms). Since they would be unresolved as separate components in COS, we have counted them as a single absorber.  We then removed any components that would be considered unmatched without the previously removed components. The results are shown in Figure~\ref{figure:number_density}b. While the number of detections per sightline slightly decreases in the $|b|$ 0-10$\degr$ bin (and $|b|$ 40-50$\degr$ bin), the $|b|$ 0-10$\degr$ bin still contains a larger number of \CIV, \SiII, and \SiIV\ absorbers than any of the other bins. This indicates that the large number of absorbers in the $|b|$ 0-10$\degr$ bin is not due entirely to the resolution differences in the STIS and COS data.

In Figure~\ref{figure:number_density}c, the number of FB HVCs remains steady with longitude, except at the highest absolute longitudes. A linear fit to the data reveals a slope of \n0.33$\pm$0.09. The data have a moderate monotonic relationship with a Spearman's rank correlation coefficient of \n0.543 and a $P$-value of 0.002. That monotonic relationship does not appear to be a linear as indicated by the low significance of the linear fit's slope.

\subsection{Velocity Centroids}
The dependence of FB HVC velocity on latitude and longitude is of high interest since it probes the velocity profile of the wind. This profile has been studied in earlier work: \citetalias{bordoloi_2017b} and \citetalias{Karim_2018} found an anti-correlation between the maximum absolute LSR velocity of the detected gas, max $|v_{\rm LSR}|$, and the absolute latitude, $|b|$. 

\begin{figure*}[!ht]
    \centering
    \epsscale{1.1}
      \plottwo{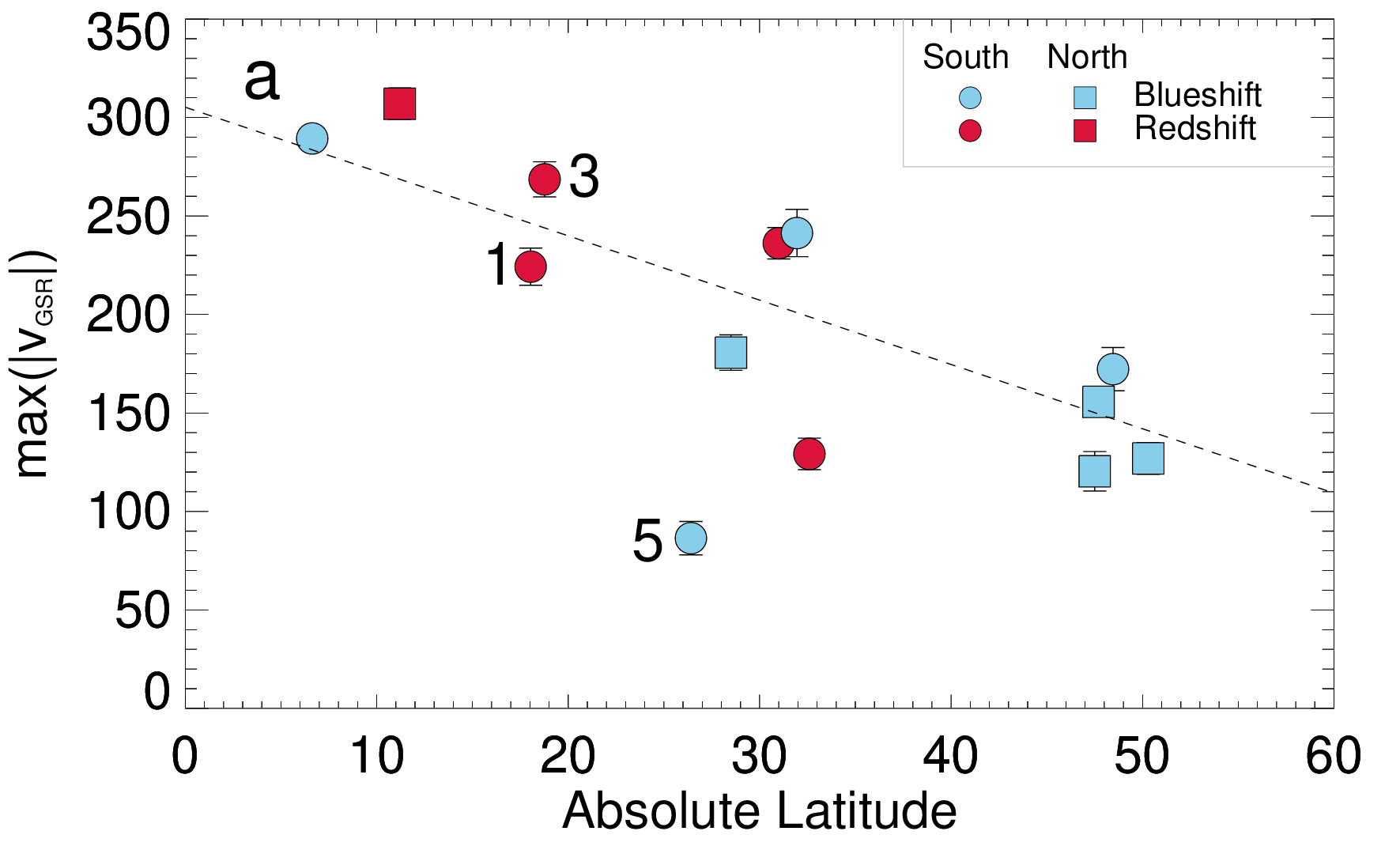}{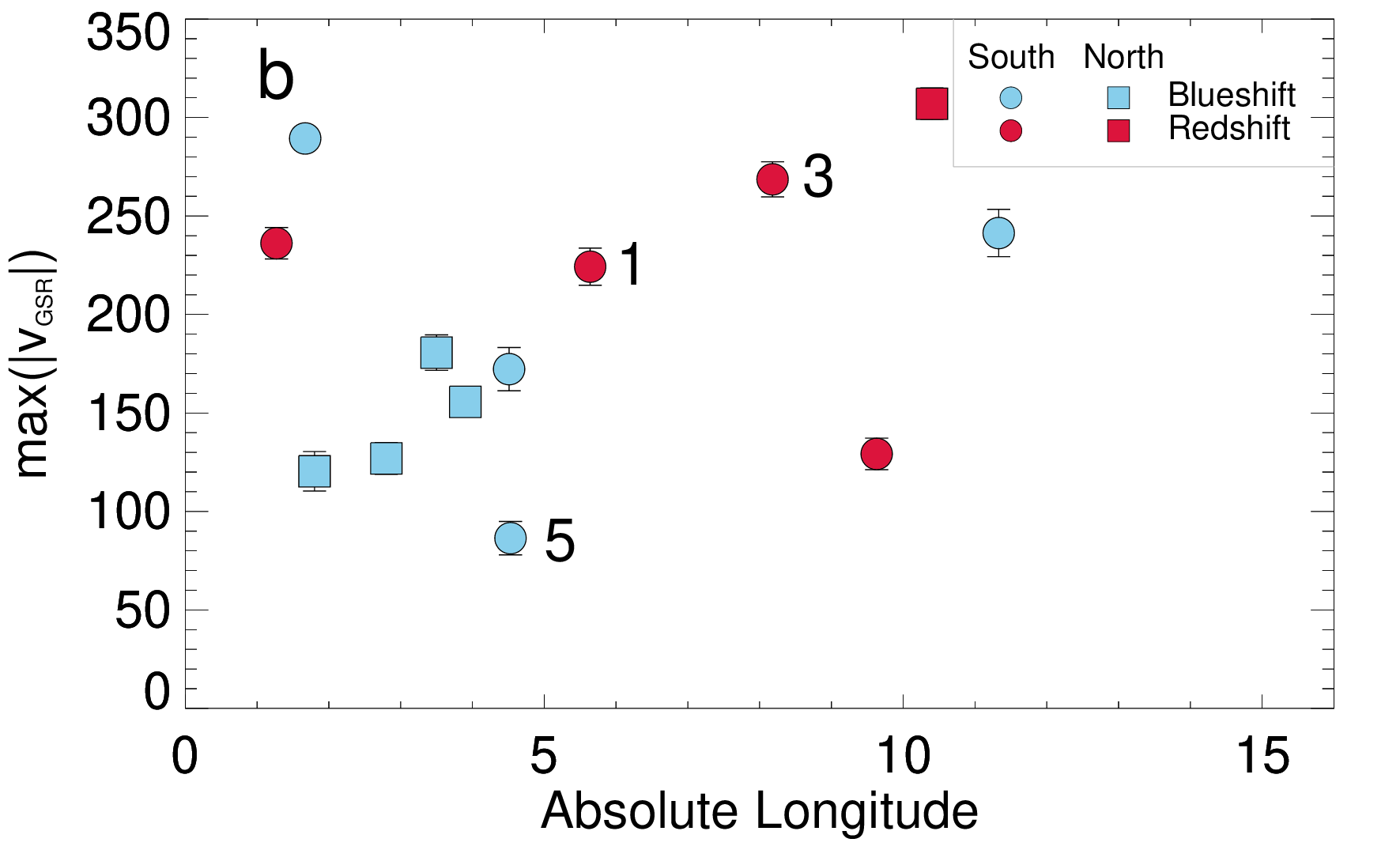}
    \epsscale{0.45}
      \plotone{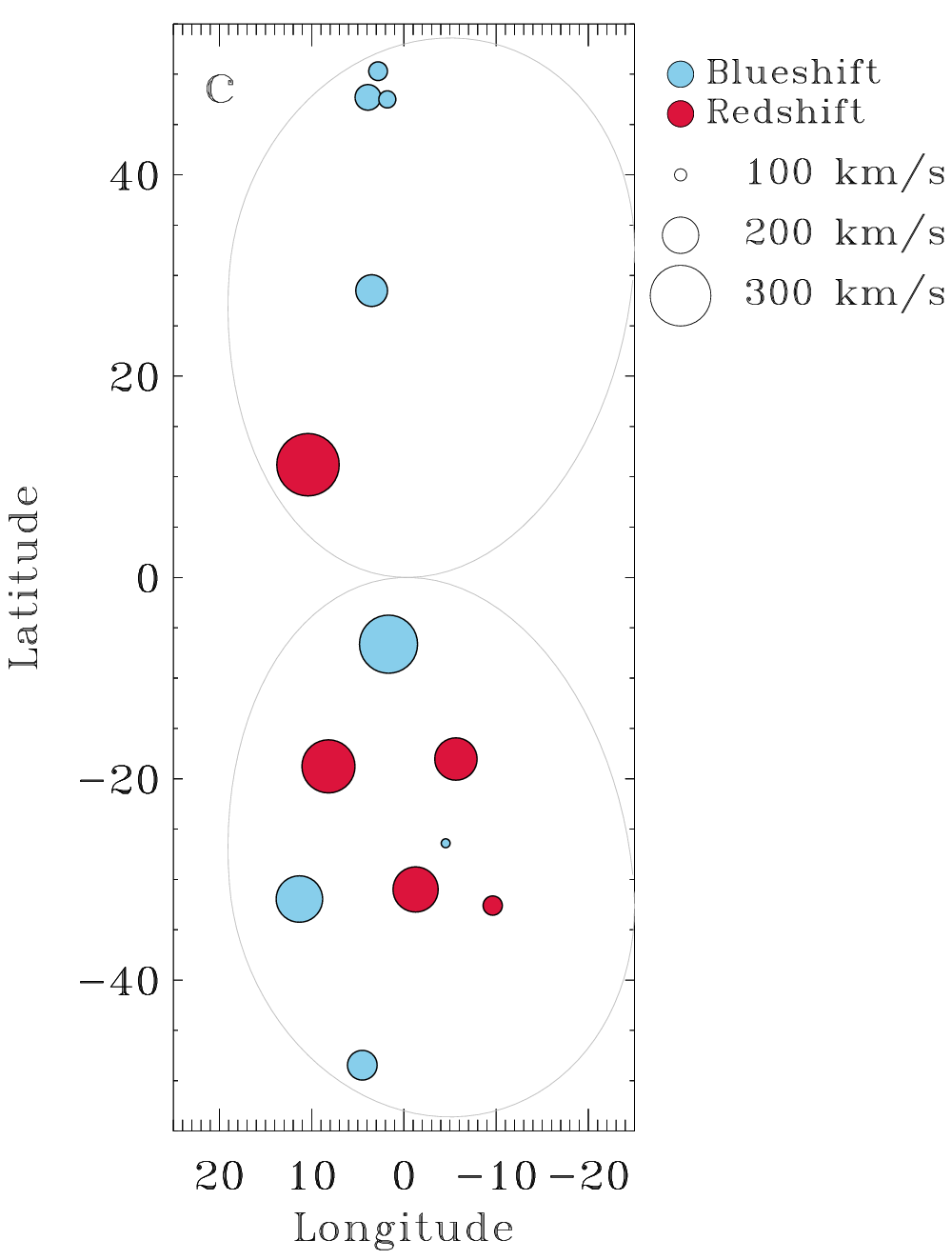}
\caption{Maximum velocity FB HVC seen as a function of location in the Bubbles for the full FB sample. In Figures (a) and (b) the squares and circles represent pointings through the northern and southern bubbles, respectively. The new low-latitude pointings are labeled in Figures (a) and (b). In all of the figures the red and blue symbols represent redshifted and blueshifted clouds, respectively. (a) Maximum GSR velocity vs. latitude. (b) Maximum GSR velocity vs. longitude. (c) Map of Fermi Bubble HVCs where the size of symbol is scaled by maximum GSR velocity. The grey ovals in (c) approximately denote the \citet{Miller_2016} FB geometry models. \label{figure:velocity}}
\end{figure*}

To determine whether the five new sightlines follow this trend and also to look for any trends in the previously unexplored axis of Galactic longitude, we plot the maximum cloud GSR absolute velocity (max$|v_{\rm GSR}|$) against $|l|$ and $|b|$ in Figures~\ref{figure:velocity}a and \ref{figure:velocity}b, where 
\begin{equation}
    v_{\text{GSR}}=v_{\text{LSR}}+(\text{cos}\,b\,\text{sin}\, l)(254\,\text{km s}^{-1}).
\end{equation}  
 We use a circular rotation velocity of 254 \kms\ at the solar distance from the Galactic center \citep{Reid_2009}. The translation to the GSR reference frame is important since it removes the effect of Galactic rotation; the latitudinal profile of the wind is largely unaffected by this translation. To calculate an FB HVC velocity we have averaged the velocity centroids of the detected ions determined to be part of the same cloud. The error bars in Figure~\ref{figure:velocity} are calculated by propagating the error of the average velocity centroids; these errors include the instrumental zeropoint error. Three sightlines in the full Fermi Bubble sample, LS 4825, PKS 2005-489, and M5-ZNG1, have STIS UV data for which a zeropoint offset of 1.7 \kms\ was used \citep{Sonnentrucker_2015}. PKS 2005-489 has one FB HVC with FUSE data for which a zeropoint offset of 10 \kms\ was used \citep{Sembach_2003}.  In Figure~\ref{figure:velocity} the blue and red symbols indicate blueshifted and redshifted clouds, respectively, while the squares and circles indicate the northern and southern Fermi Bubbles, respectively.

We find that $|v_{\rm GSR}|$ decreases with increasing latitude as shown in Figure~\ref{figure:velocity}a. Linear fits to this trend yields:
\begin{equation}
     {\rm max}(|v_{\rm GSR}|)= (-3.3\pm0.1)b+(305.3\pm3.1){\rm km\,s}^{-1}
\end{equation}
where $b$ is in degrees.

The dependence of maximum GSR velocity on Galactic longitude is shown in Figure~\ref{figure:velocity}b. No clear monotonic trends are found in the current data set ($\rho_{\text{S}}=0.225$ and $P$-value=0.459).

To study the trends of velocity in both Fermi Bubbles separately and to better visualize any trends and symmetries, we have displayed a map of the maximum absolute GSR velocities in Figures~\ref{figure:velocity}c.  In this map, the size of the circle is scaled with velocity and the color of the circle represents redshifted/blueshifted components. In Figure~\ref{figure:velocity}c the velocities decrease with increasing latitude, which confirms that general trend seen in Figure~\ref{figure:velocity}a. We do not see any obvious asymmetries in the maximum velocity of clouds in the northern and southern Fermi Bubbles.

\subsection{Component Kinematics}
As part of this study, we analyzed the kinematics of the Fermi Bubble HVC population. We focus on two separate kinematic measurements: the velocity centroid alignment between different ions, and the distribution of component linewidths ($b$-values) for different ions. Both of these measurements help us to determine which absorption components are part of the same FB HVC and to reveal the phase structure of the clouds, i.e. whether they are single-phase or multi-phase. The velocity offsets may also reveal the presence of any evaporative or condensing flows \citep{Borkowski_1990}. For example, gas in collisional ionization equilibrium would result in lines that all have the same centroid velocity and line width, independent of whether they are high ions (\CIV\ and \SiIV) or low ions (\CII\ and \SiII). In contrast, a cool gas with an evaporative transition layer may give rise to velocity centroid offsets between the low and high ions. For a review of ionization mechanisms in HVCs see \citet{Fox_2004}. Below we present the quantitative results of the velocity centroid matching and $b$-value analysis.
  
\subsubsection{Velocity Centroid Offsets}
We measured the difference between the velocity centroids ($\Delta v$) of different ions in each FB HVC. Using only the matched FB HVC components (velocity centroid offset $\le3\sigma$), we measured the mean 
$\Delta v$, the standard error of the mean, and the standard deviation of the mean for high-low, high-mid, mid-low, high-high, and low-low ion pairs (e.g. \CIV-\CII, \SiIV-\SiIII, \SiIII-\SiII, \SiIV-\CIV, and \SiII-\CII\ pairs, respectively). The standard error of the mean is a measure of the accuracy of the mean calculation when compared to the parametric mean and is equal to the standard deviation divided by the square root of the number of pairs. The 3$\sigma$ cutoff imposed in our matching process does not bias these calculations towards low $\Delta v$ values compared to visual matching techniques, as discussed in Appendix~\ref{appendix:centroid_offset}. We list the results in Table~\ref{table:subtraction}. 


\begin{deluxetable}{lcccc}[!ht]
\tablecaption{Velocity Centroid Offsets \label{table:subtraction}}
\tablehead{\colhead{Ion Pair} & \colhead{$\langle\Delta v \rangle$} & \colhead{Standard Error} & \colhead{Standard} & \colhead{Number of}\\[-8pt]
\colhead{} & & \colhead{of the Mean}& \colhead{Deviation} &\colhead{Pairs}\\[-8pt]
 & \colhead{(\kms)} & \colhead{(\kms)} & \colhead{(\kms)} &}
\startdata
High-Low & 9.3 & 1.2 & 9.4 & 64\\
High-Mid & 7.8 & 1.7 & 7.8 & 22\\
Mid-Low & 8.0 & 1.1 & 7.0 & 43\\
Low-Low & 4.0 & 0.5 & 5.5 & 146\\
High-High & 11.0 & 3.2 & 12.6 & 16\\\hline
\enddata
\tablecomments{This table shows the mean velocity centroid offset, its standard error, and the standard deviation of $\Delta v$ for various ion pairs for all matched components in the full FB sample (including absorption features matched to some, but not all the lines in a given cloud).}
\end{deluxetable}

\subsubsection{Component Line Widths}

The velocity width ($b$-value) of individual ion absorption components encodes information on the thermal state of the gas. It contains thermal ($b_{\rm therm}$) and non-thermal contributions ($b_{\rm non}$), and can be generally written as 
$b^2=b^2_{\rm therm}+b_{\rm non}^2=2kT/Am_{\rm H}+b_{\rm non}^2$,
where $A$ is the atomic weight of the ion.

\begin{figure}[!ht]
\centering
\epsscale{1.1}
\plotone{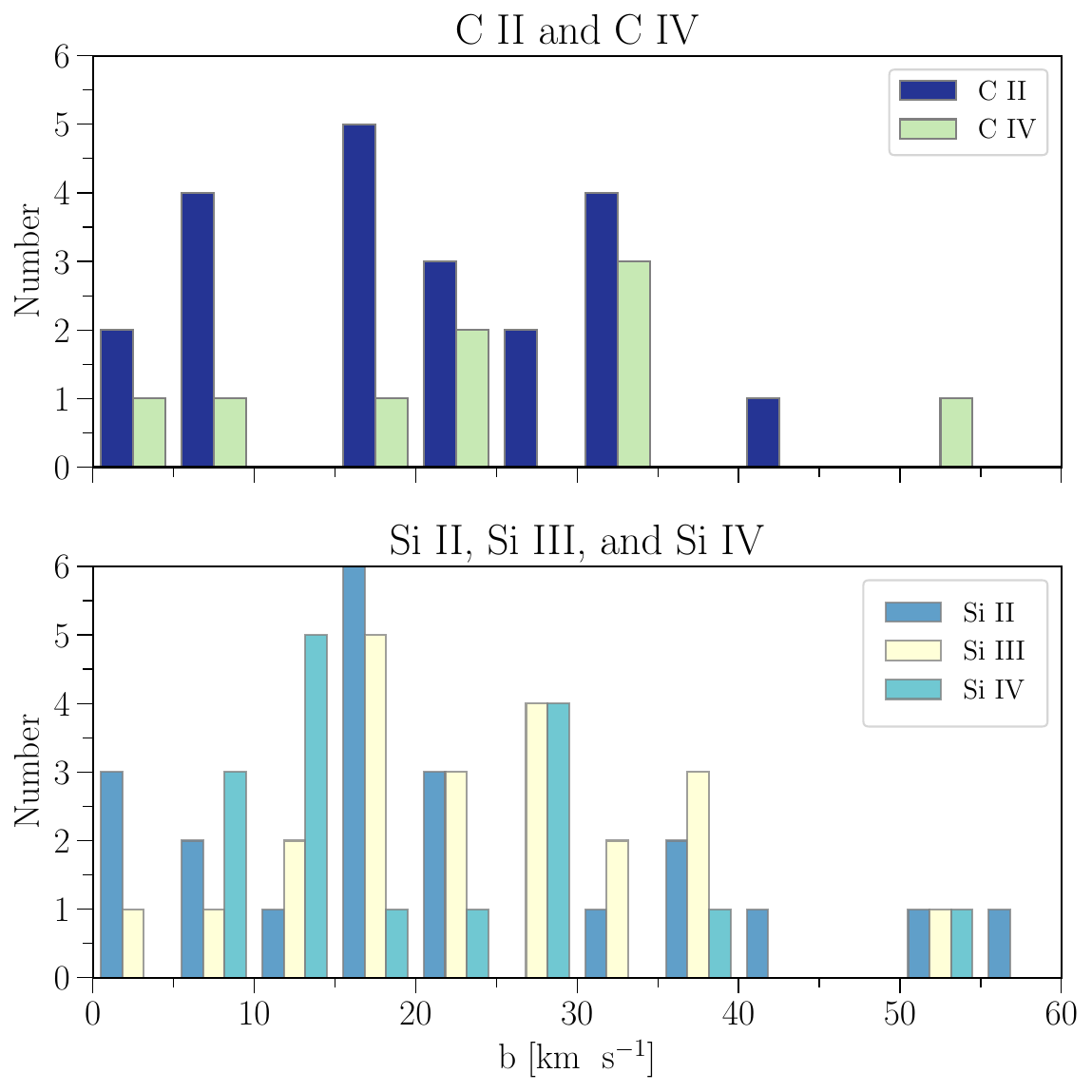}
\caption{Distribution of $b$-values for Fermi Bubble HVCs in the full FB sample. The top panel compares \CII\ and \CIV\ and the bottom panel compares \SiII, \SiII, and \SiIV. There is no statistical evidence for any differences between these distributions, as shown in Table~\ref{table:K-Stest}.} 
\label{figure:b_values}
\end{figure}

In Figure~\ref{figure:b_values} we compare the $b$-value distributions of the Fermi Bubble HVCs for \CII\ and \CIV\ (top panel) and for \SiII, \SiII, and \SiIV\ (lower panel). We only compare ions of the same species since thermal broadening is dependent on atomic weight. We combine all of our FB HVC data in this plot in order to analyze the general trends seen in the FB HVCs in a statistically significant manner. The mean $b$-values for silicon and carbon are 22.4 and 21.1, respectively.

Two-sided K-S tests were run to check for the statistical similarity of the $b$-value distributions of different ions. The resulting $P$-values and $D$-statistics are shown in Table~\ref{table:K-Stest}. None of the K-S tests provide any evidence that the $b$-values are drawn from separate populations. Since we have not analyzed each FB HVC component separately, we cannot conclude with certainty that the high and low ion $b$-values for every FB HVC trace each other well. For example, PDS 456 and M5-ZNG1 each have two FB HVCs with high ion absorbers that are much broader than the low ion absorbers. Yet, in general the trends in the $b$-values indicate that the high and low ion $b$-values are similar over the entire sample of FB HVCs. This indicates that interpreting the $b$-value distribution is not simple. 

\begin{deluxetable}{ccc}[!ht]
\tablecaption{K-S Tests for $b$-values}
\tablehead{\colhead{Ions} & \colhead{$D$} & \colhead{$P$-value}}
\startdata
\SiII\ and \SiIV\   &  0.238  & 0.619  \\
\SiII\ and \SiIII\  &  0.208  & 0.689 \\
\SiIII\ and \SiIV\  &  0.318  & 0.251   \\
\CII\ and \CIV\     &  0.286  & 0.604   \\
\enddata
\tablecomments{Two-sided KS tests were computed for each pair of ions shown for the data in Figure~\ref{figure:b_values}.}
\label{table:K-Stest}
\end{deluxetable}

\subsection{Ion Column Densities}
Another key property of the Fermi Bubble HVC population is the dependence of column density on position, which may indicate changes in physical conditions of the clouds as they rise into the halo. 

\begin{figure*}[ht!]
    \centering
    \epsscale{0.92}
      \plottwo{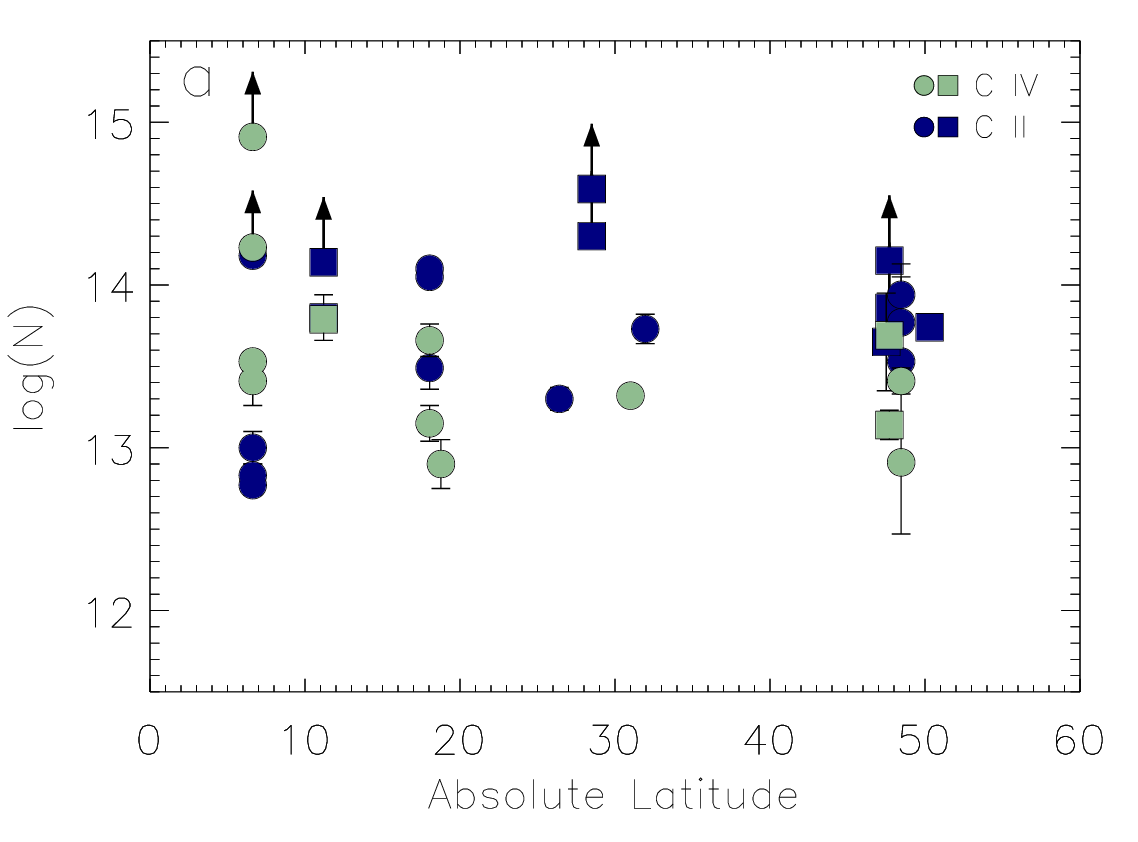}{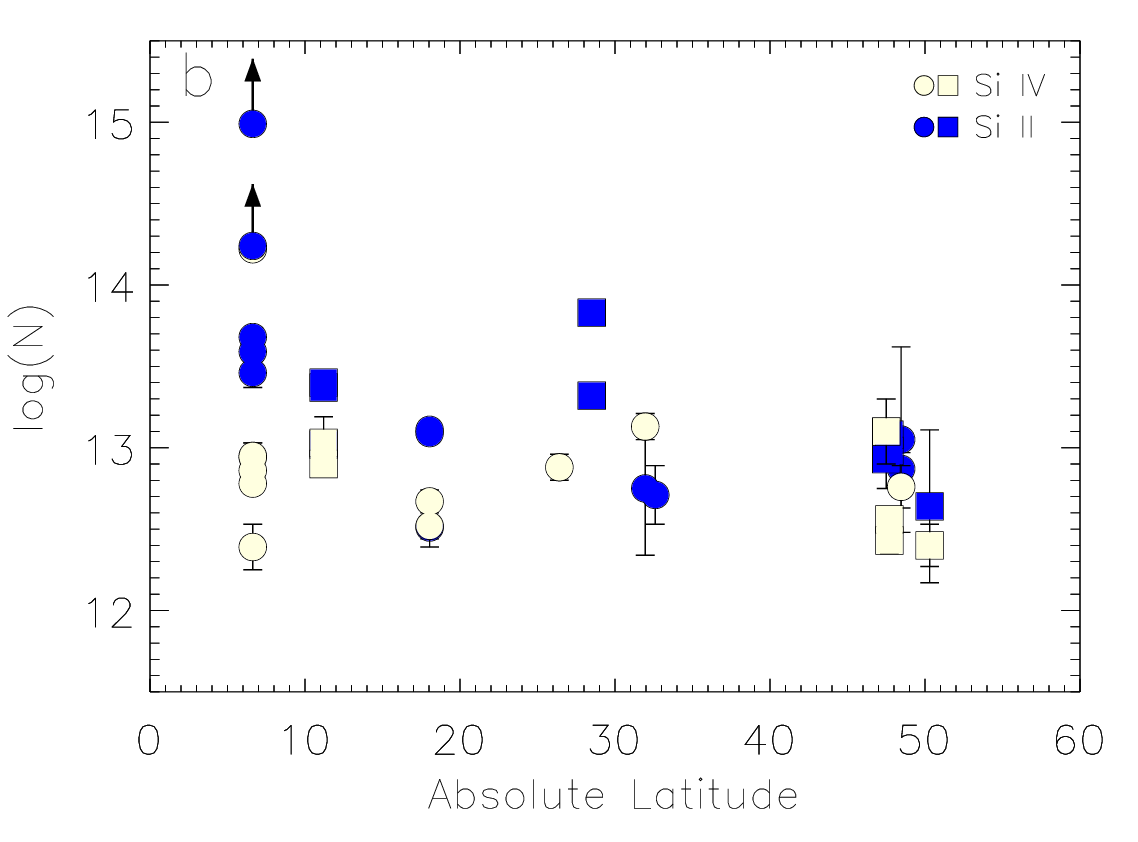}
      \plottwo{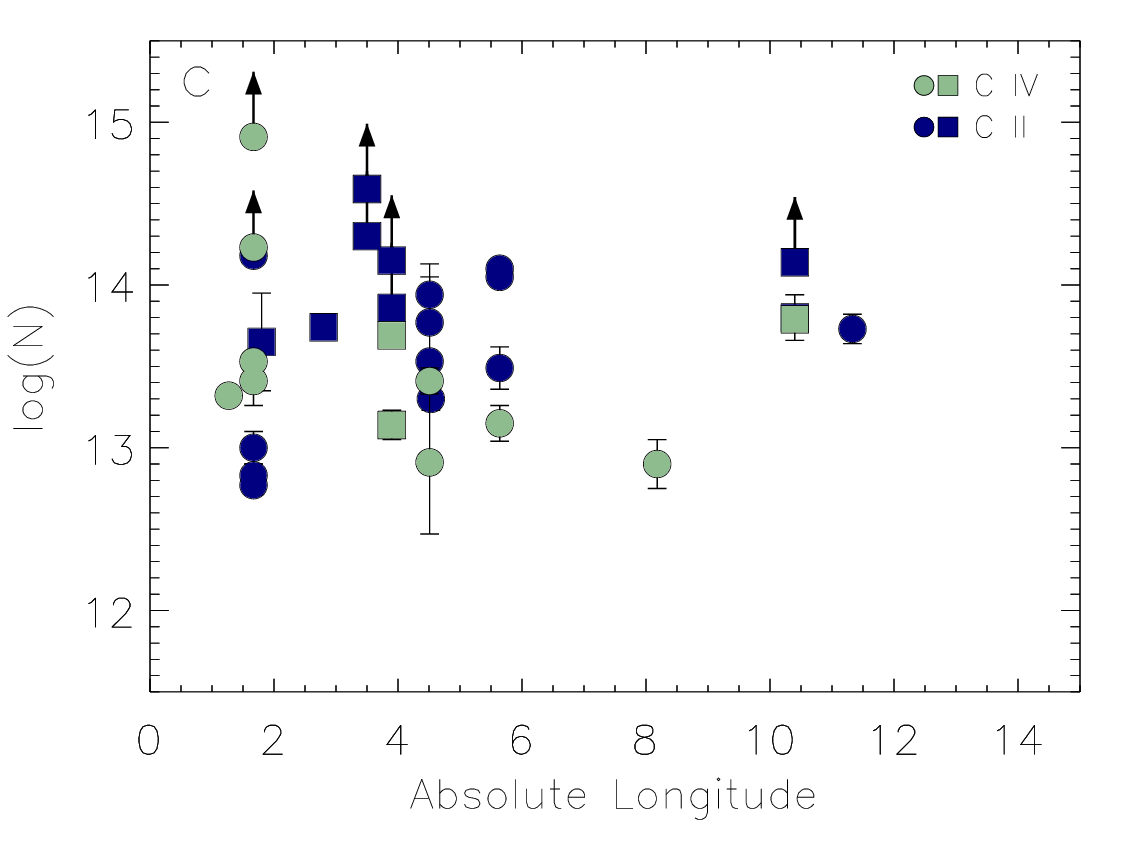}{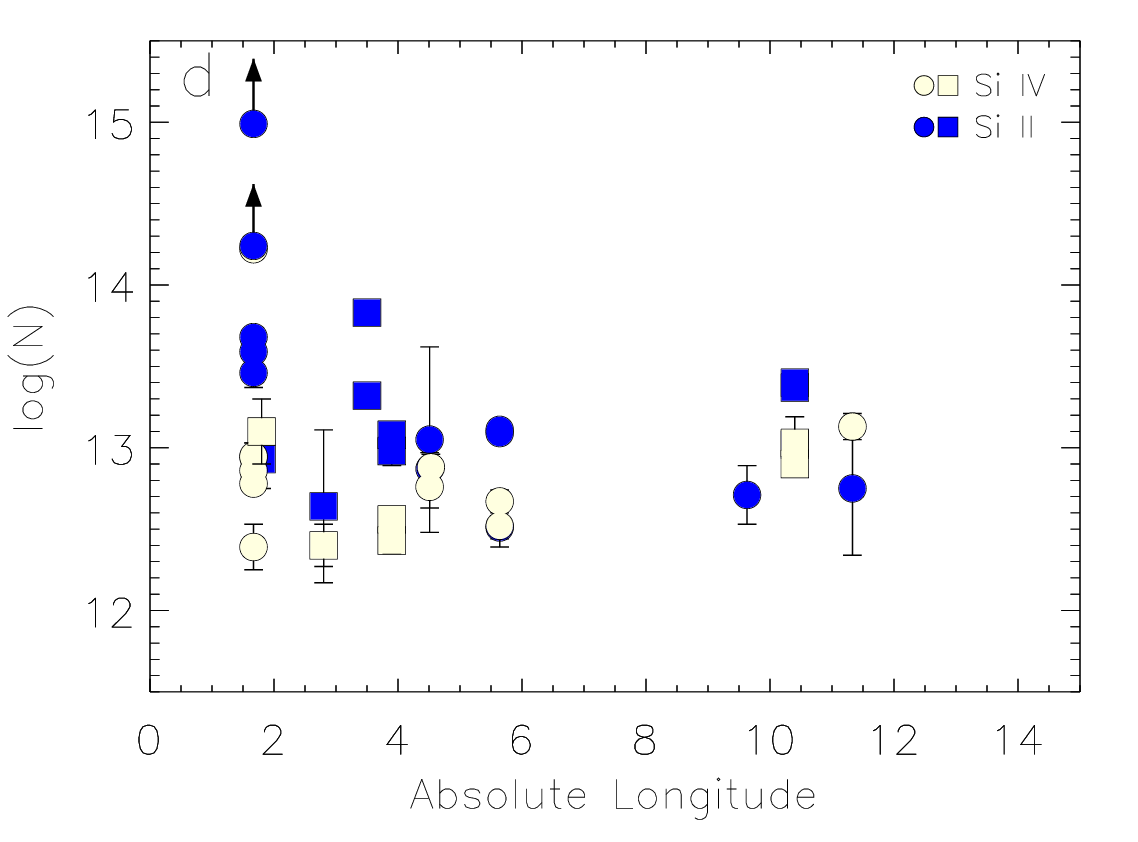}
      \caption{Plots of the ionic column density as a function of position for each FB HVC.  The squares and circles represent pointings through the northern and southern bubbles, respectively. (a-b) The individual FB HVC column densities \frenchspacing{vs. }latitude. (c-d) The individual FB HVC column densities \frenchspacing{vs. }longitude. \label{figure:column}}
\end{figure*}

We have plotted the carbon and silicon ion column densities from all FB HVCs detected in the full FB sample and plotted them against Galactic latitude and longitude in Figure~\ref{figure:column}. C and Si are useful elements to trace since we can measure columns for each element in both high and low ions. We modeled a linear fit to each ion separately using the \texttt{survreg} function in \texttt{survival}\footnote{https://CRAN.R-project.org/package=survival} package from \textsf{R}\footnote{https://www.R-project.org/} \citep{Therneau_2000}. The \texttt{survival} package fits a parametric survival regression model, allowing the lower limits to be included in the linear fit model (errors for other measurements are not included in the model). For the calculations, we set the maximum number of iterations to be 100 and the assumed distribution of $log(N)$ to be Gaussian. The results are listed in Table~\ref{table:column_fits}. The standard error listed for each coefficient is a measure of coefficient standard deviation, calculated by taking the square root of the variance-covariance matrix diagonal elements. The $P$-value is also listed for the ``A'' coefficient (all $P$-values for intercepts were reported as \textless$2*10^{-16}$). The results of the linear modeling indicate that none of graphs in Figure~\ref{figure:column} have a slope significantly greater than zero. We therefore conclude that there is no evidence in the current data for a change in the ion columns with latitude or longitude.



\begin{deluxetable*}{lccccccccc}[!ht]
\tablecaption{Dependence of Column Density on Latitude and Longitude\label{table:column_fits}}
\tablehead{\colhead{} & \colhead{Equation} & \colhead{Element} & \colhead{A} & \colhead{Standard Error} & \colhead{$P$-value} & \colhead{B} & \colhead{Standard Error}} %
\startdata
\multirow{4}{*}{Galactic Latitude} & \multirow{4}{*}{log\,$N$=A$b$+B}  
& \CII\ & 0.0093 & 0.0083 & 0.262 & 13.60 & 0.27 \\ 
& & \CIV\ & $-$0.0159 & 0.0081 & 0.050 & 13.94 & 0.24  \\ \cline{3-8}   
& & \SiII\ & $-$0.0199 & 0.0064 & 0.002 & 13.78 & 0.19  \\
& & \SiIV\  & $-$0.0018 & 0.00572 & 0.160 & 13.05 & 0.17 \\\hline 
\multirow{4}{*}{Galactic Longitude} & \multirow{4}{*}{log\,$N$=A$l$+B} &
\CII\ & 0.0495 & 0.0491 & 0.313 & 13.63 & 0.27  \\
& & \CIV\  & \n0.0672 & 0.0565 & 0.234 & 13.84 & 0.28  \\ \cline{3-8} 
&  & \SiII\ & $-$0.0740 & 0.036 & 0.038 & 13.64 & 0.21  \\
& & \SiIV\  & 0.0004 & 0.0329 & 0.990 & 12.86 & 0.18  \\
\enddata
\tablecomments{This table contains parameters from linear fit models to the ionic column densities vs. latitude and longitude using parametric survival regression. The coefficients each have a standard error listed and the $P$-values of the slope are also given (all $P$-values for B were reported as \textless$2*10^{-16}$).}
\end{deluxetable*}

\begin{figure*}[ht!]
    \centering
    \epsscale{0.377}
      \plotone{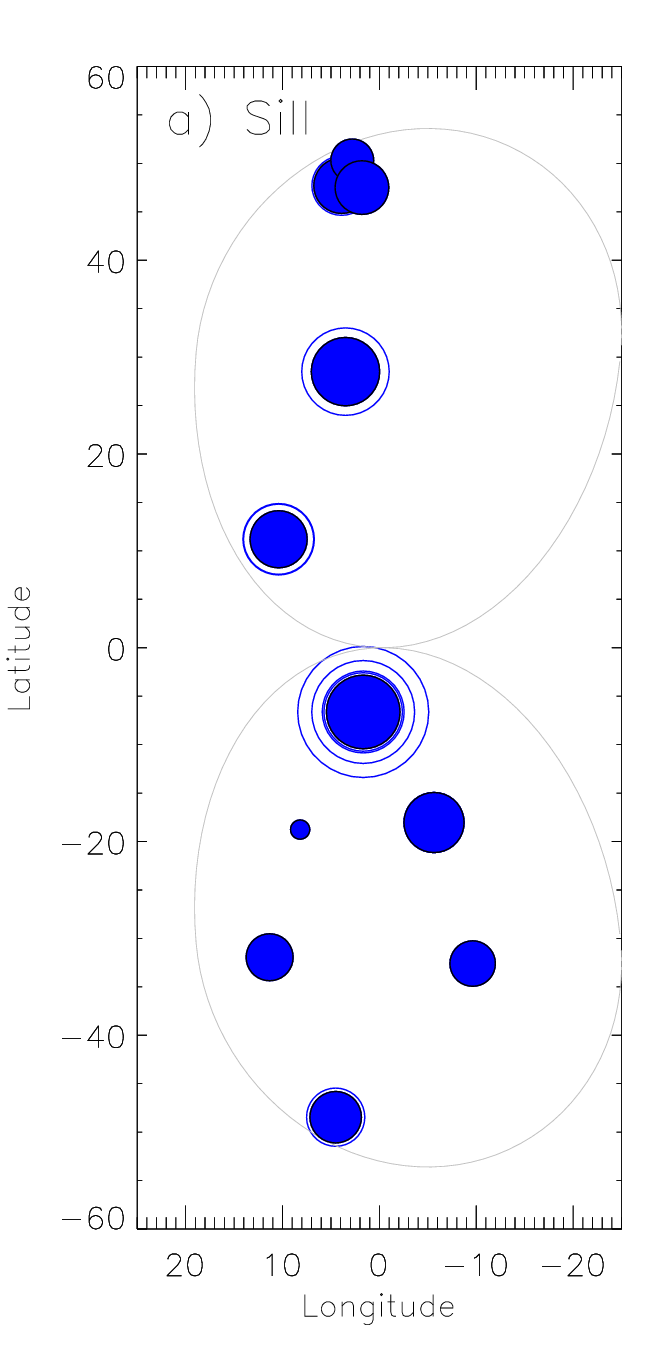}
    \epsscale{0.5826}
      \plotone{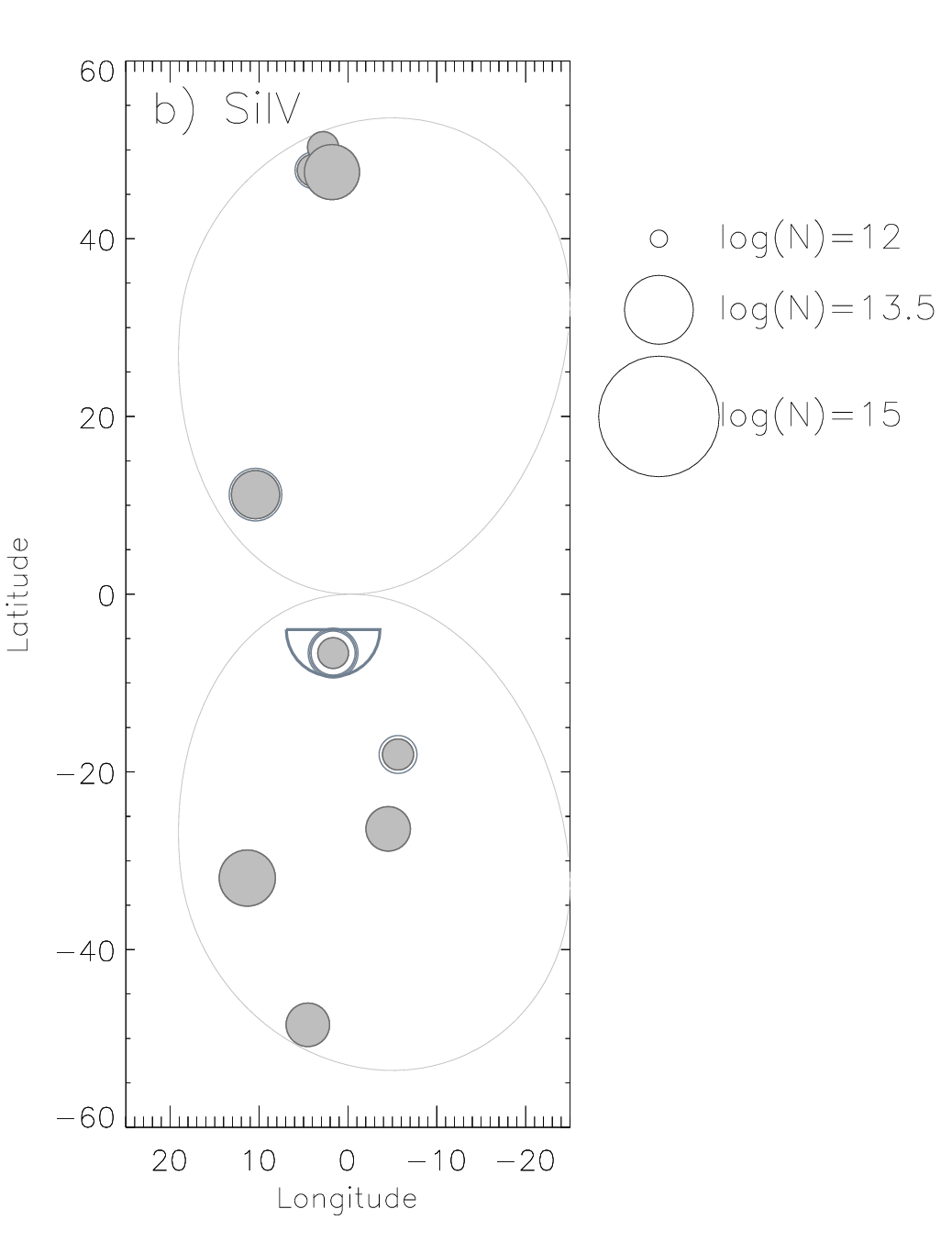}\\
    \epsscale{0.377}
      \plotone{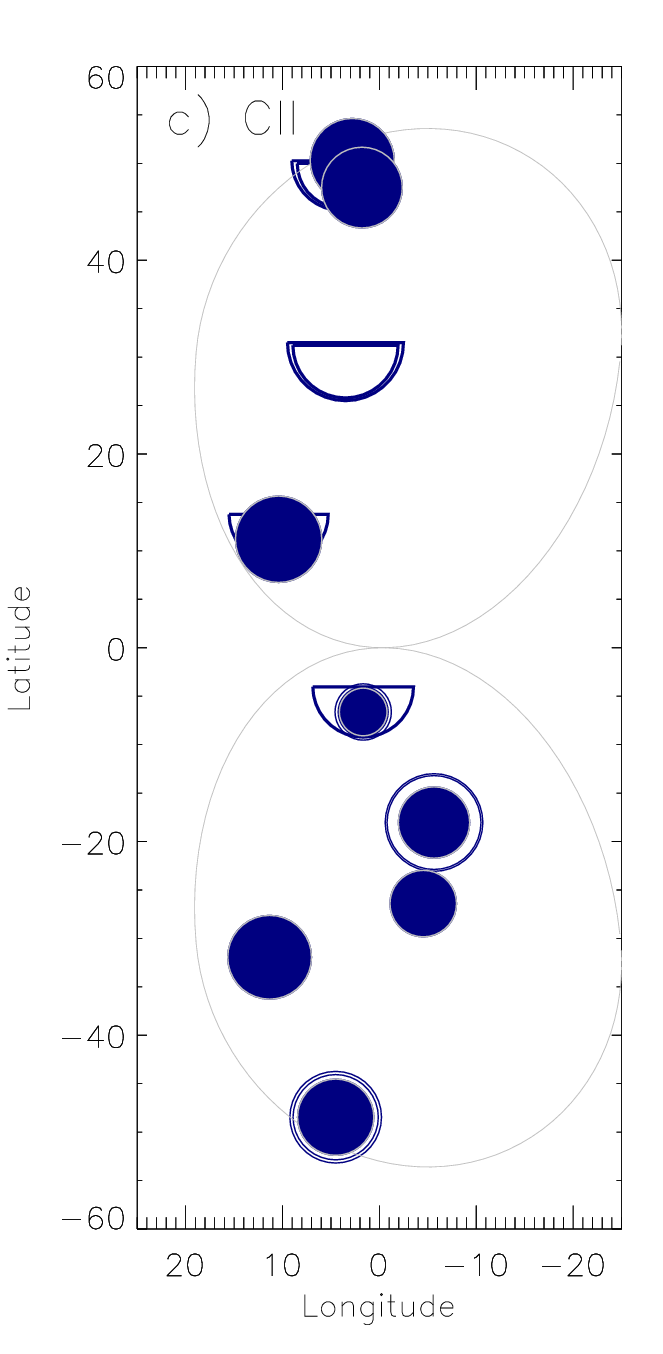}
    \epsscale{0.5826}
      \plotone{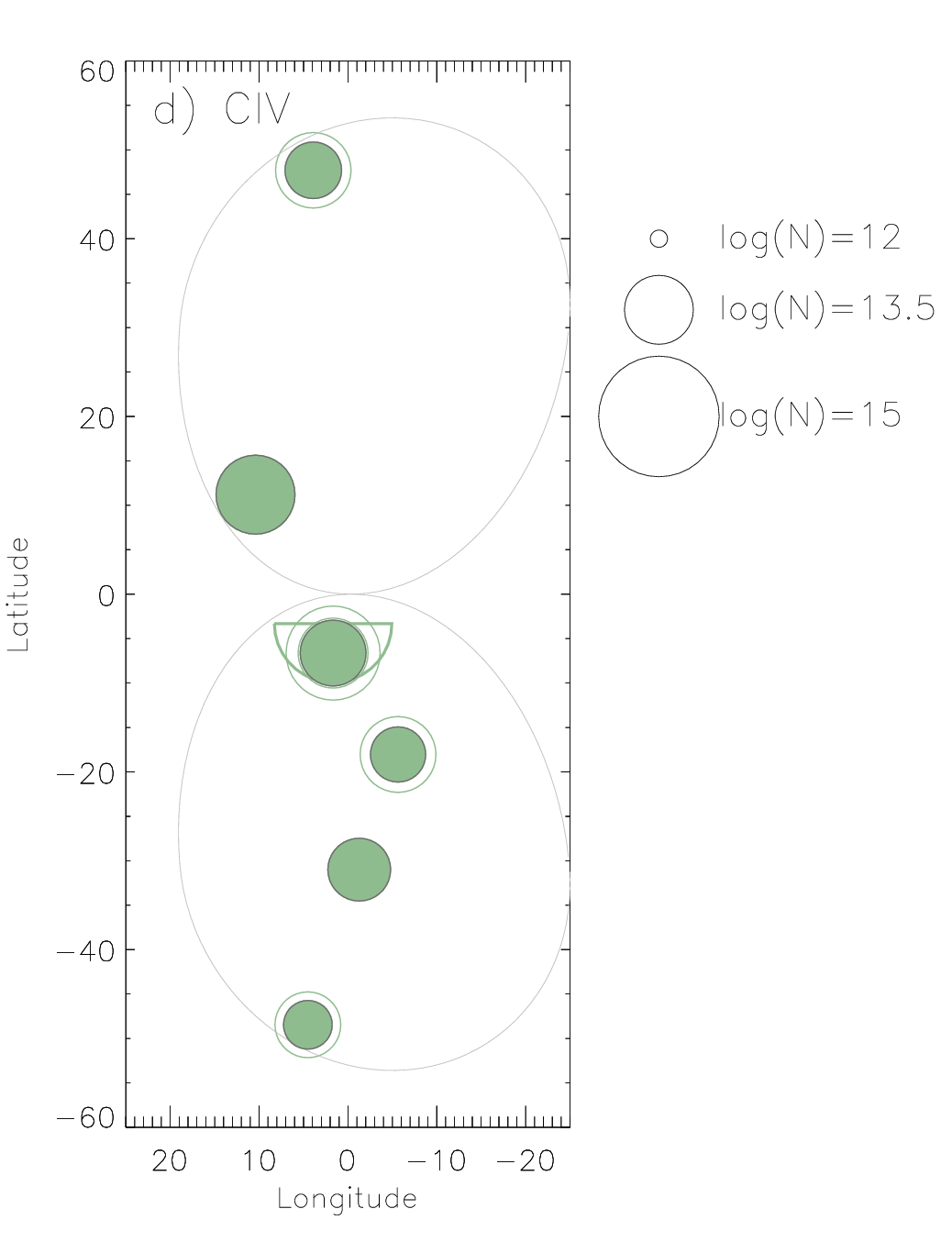}
      \caption{Plots of the ionic column density as a function of position for each FB HVC. Filled circles are the lowest column measurement for each pointing. The half circles represent lower limits. The size of each circle is scaled to column density, as shown in the key between plots denoted by black circles. The grey ovals denote the \citet{Miller_2016} FB models. \label{figure:column_full}}
\end{figure*}

To study the column densities in both Fermi Bubbles separately and to better visualize any potential trends and symmetries, we have plotted maps of \CII, \CIV, \SiII, and \SiIV\ column densities for each FB HVC in Figure~\ref{figure:column_full}.  In these maps, the size of the circle is scaled with log(N) with the lowest column for each pointing indicated by a filled circle. No new trends or asymmetries are obvious in these plots.

\subsection{Ion Ratios}\label{section:ion_ratios}

The ionization level is another property of interest in the Fermi Bubble cloud population. This property can be diagnosed using the following UV line ratios: \CIV/\CII, \SiIV/\SiII, \SiIV/\SiIII, and \SiIII/\SiII. Each of these ratios compares two ions from the same element, thus ensuring that they are insensitive to dust depletion and absolute abundance variations. In previous work, \citetalias{Karim_2018} did not find a trend in silicon ionization levels with latitude in their data when combined with the data of \citetalias{Fox_2015}, \citetalias{bordoloi_2017b}, and \citetalias{Savage_2017}.

\begin{figure*}[ht!]
    \centering
    \epsscale{1}
    \plottwo{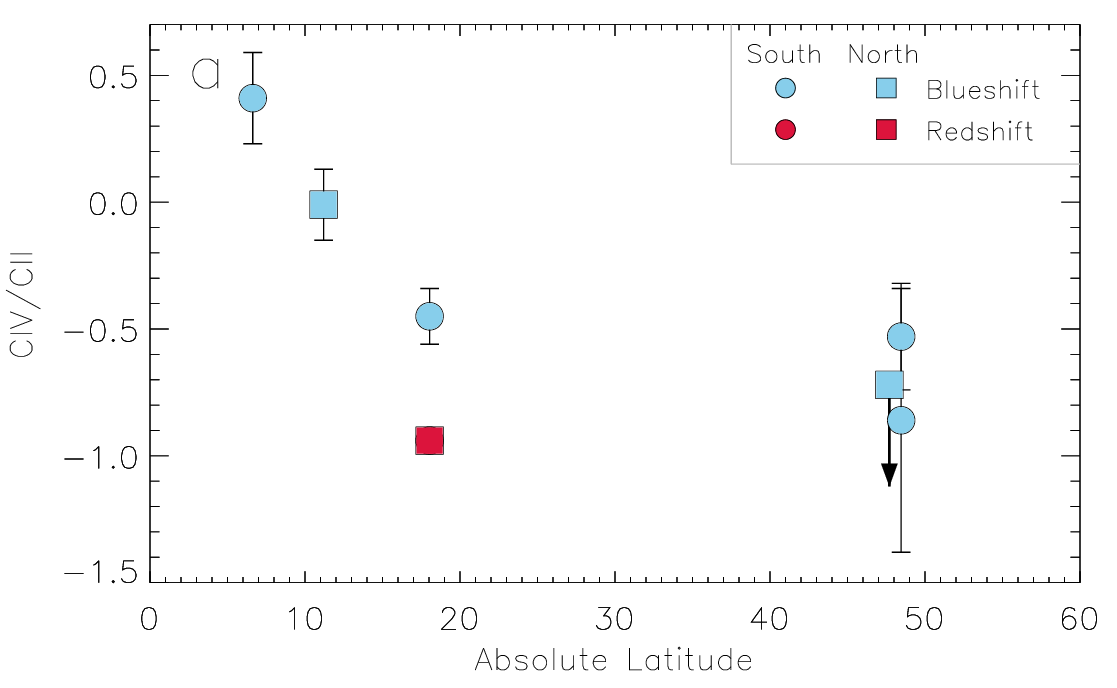}{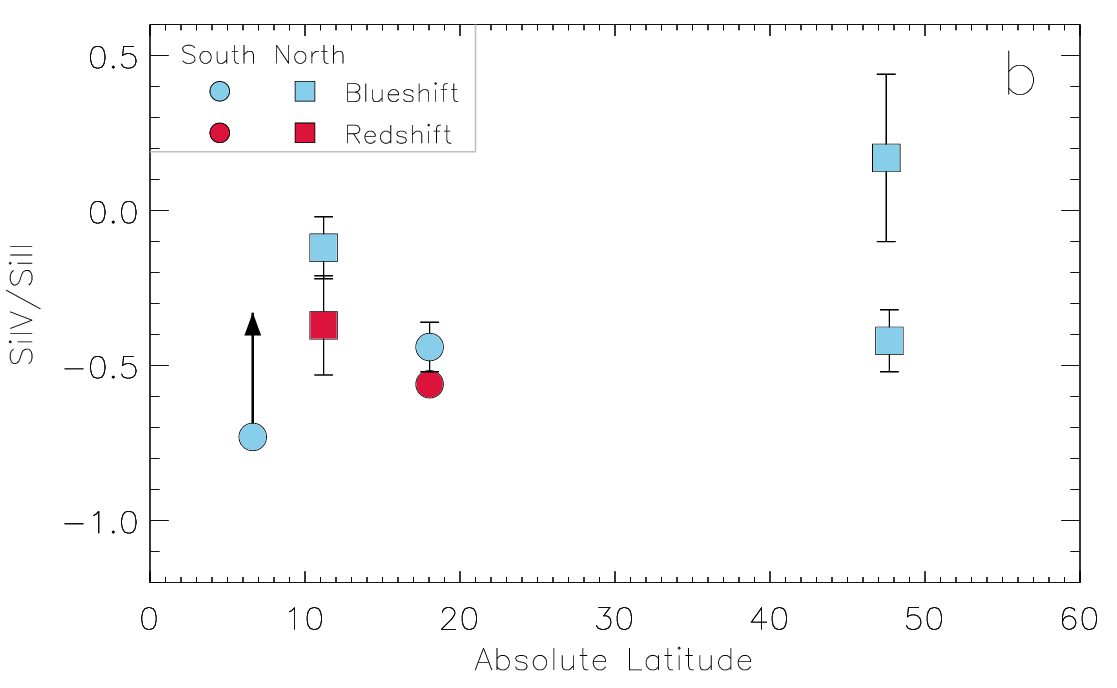}
    \plottwo{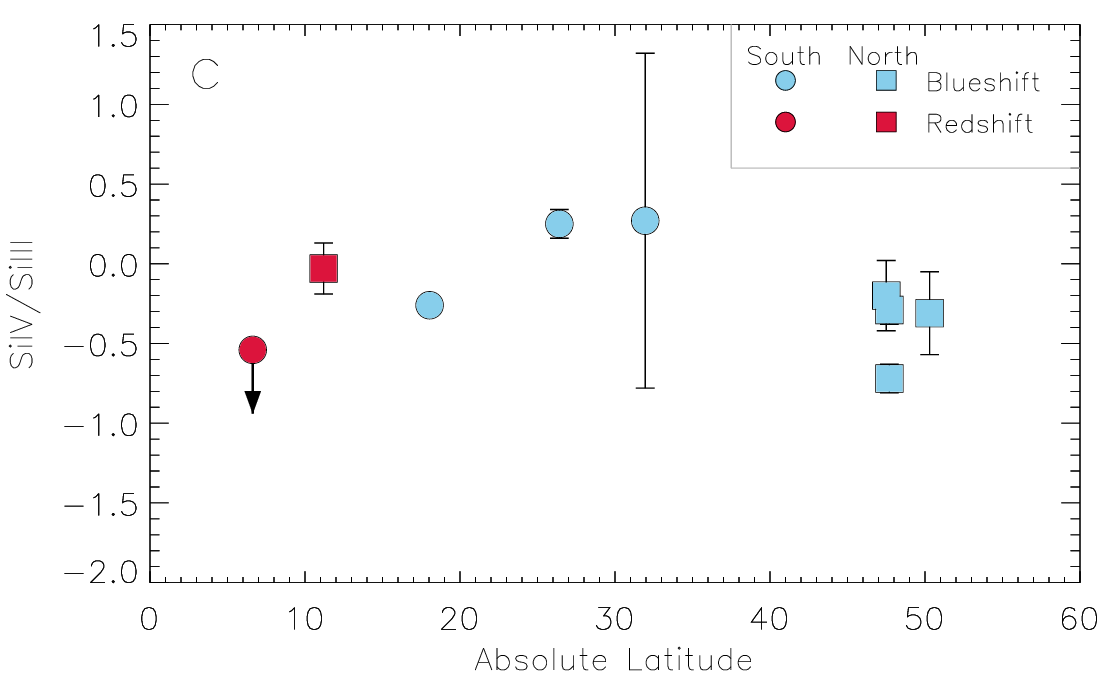}{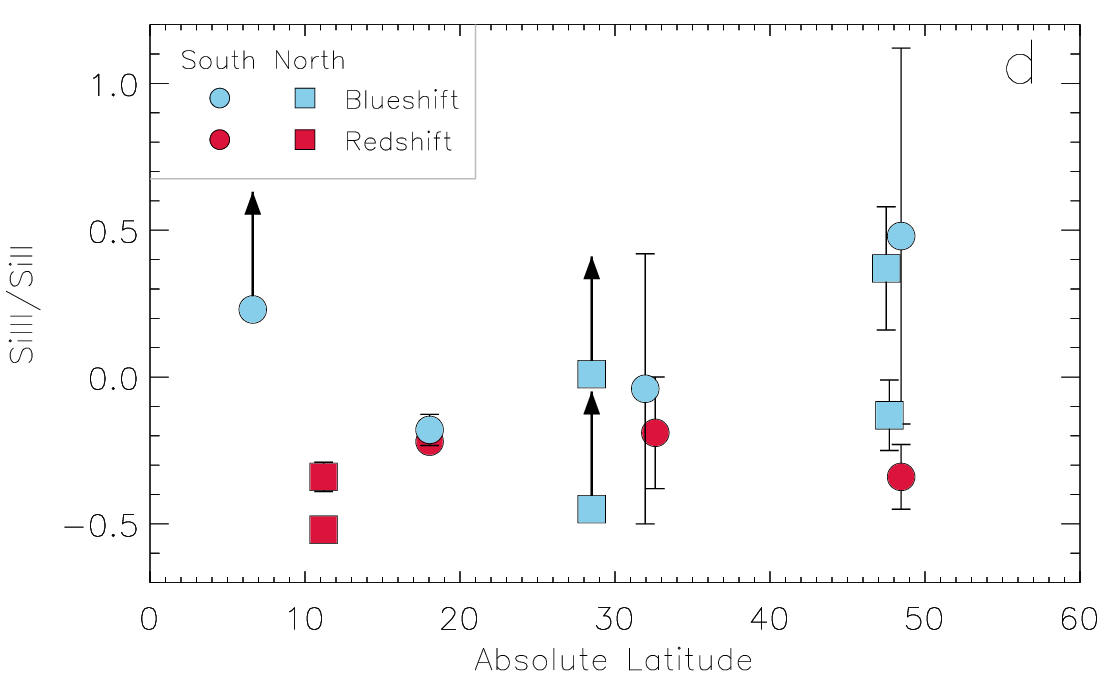}
    \plottwo{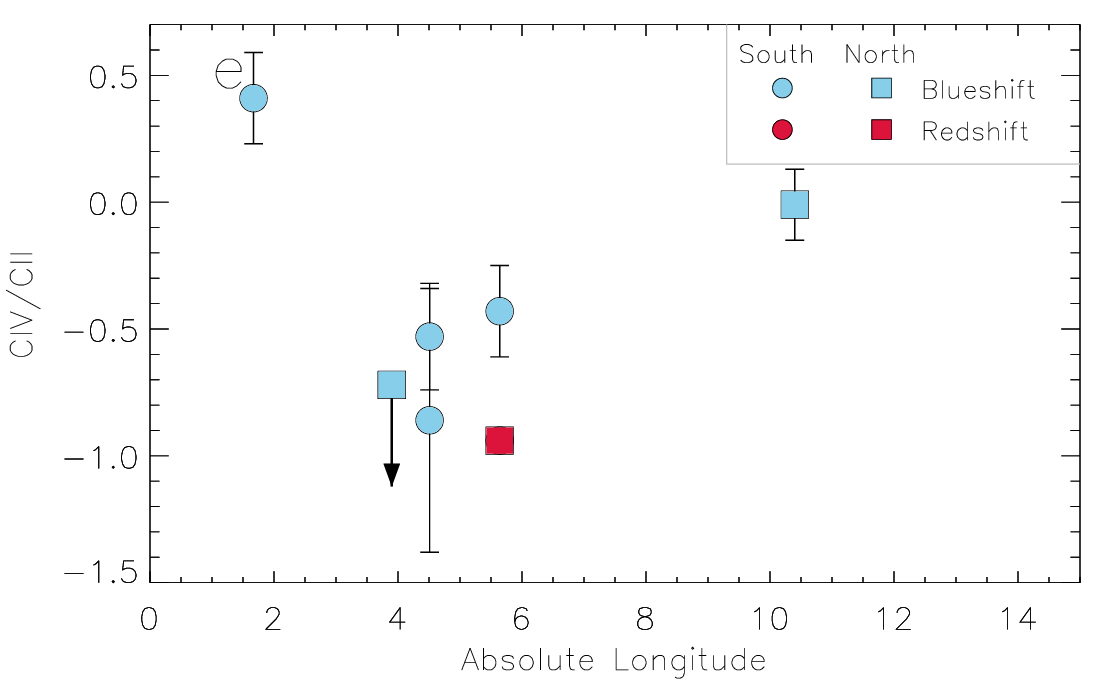}{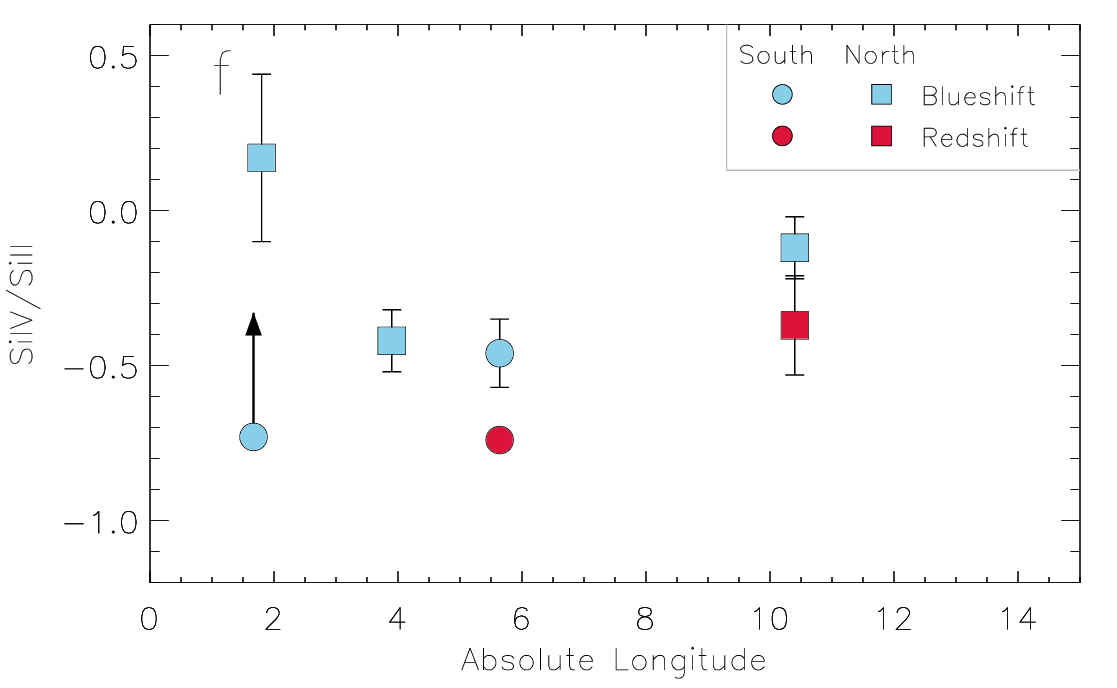}
    \plottwo{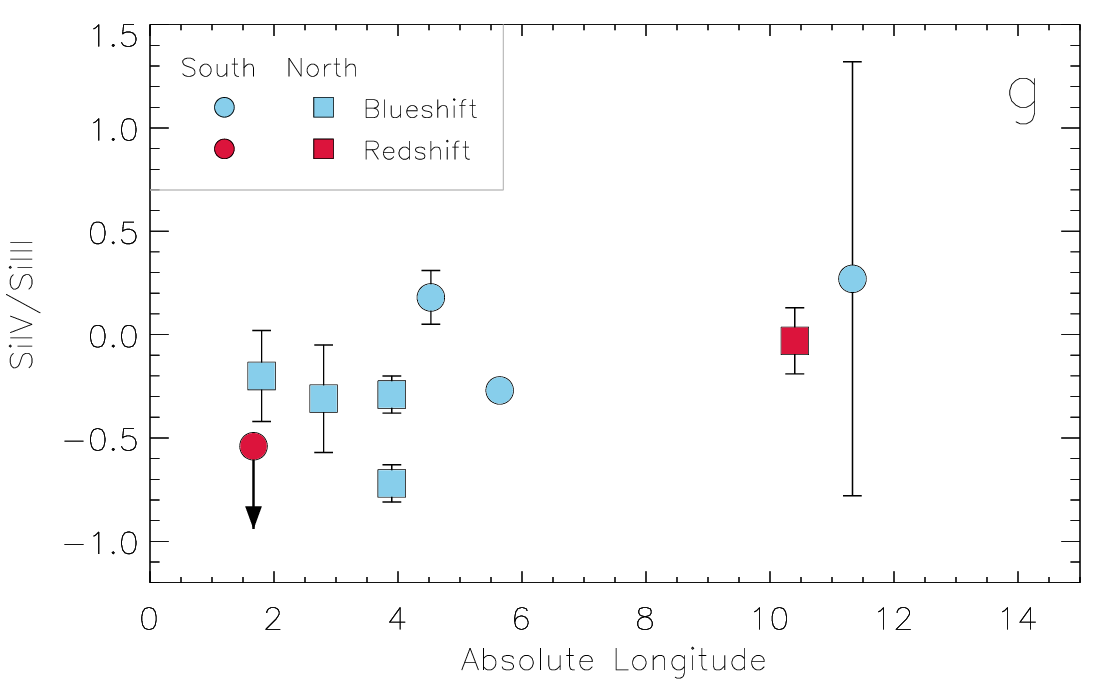}{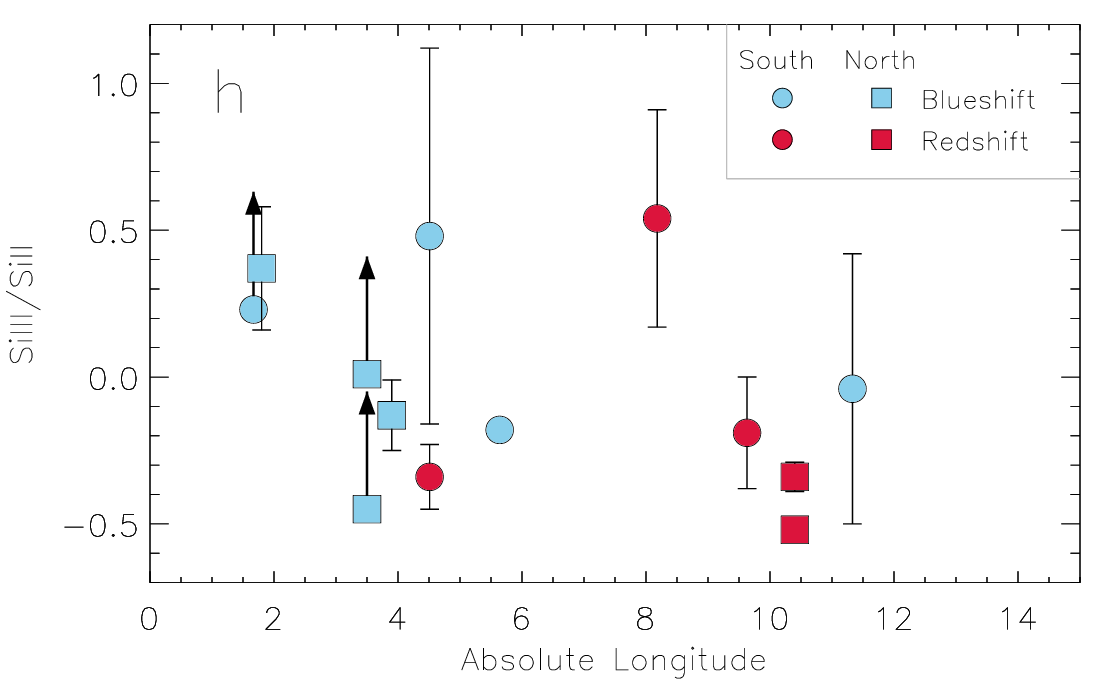}
\caption{Log of the column density ratios of Fermi Bubble HVCs as a function of latitude and longitude.}
\label{figure:ion_ratios}
\end{figure*}

Figure~\ref{figure:ion_ratios} displays the ion ratios for each FB HVC from the full FB sample and compares them with Galactic latitude and Galactic longitude.  We modeled a linear fit to each ratio separately using the \texttt{survival} package in \textsf{R} for graphs containing lower limits and using a combination of the \texttt{survival} and \texttt{NADA}\footnote{https://CRAN.R-project.org/package=NADA} packages in \textsf{R} for graphs containing upper limits. The \texttt{NADA} package handles upper limits to be included in the linear fit model from the \texttt{survival} package. For the calculations, we set the maximum number of iterations to be 100 and the assumed distribution of $log(N)$ to be Gaussian. The results are listed in Table~~\ref{table:ratio_fits}. None of the slopes significantly deviate from zero; the current data does not indicate that the ion ratios are changing with latitude or longitude. 



 \begin{deluxetable*}{cccccccc}[t]
\tablecaption{Linear Fits to Ion Ratios vs. Latitude and Longitude \label{table:ratio_fits}}
\tablehead{\colhead{} & \colhead{Equation} & \colhead{A} & \colhead{Standard Error} & \colhead{$P$-value} &\colhead{B} & \colhead{Standard Error} & \colhead{$P$-value}} 
\startdata
 \multirow{4}{*}{Galactic Latitude} &
 \multirow{4}{*}{$\left.\begin{array}{l} 
 \text{\CIV/\CII} \\
 \text{\SiIV/\SiII} \\
 \text{\SiIV/\SiIII} \\
 \text{\SiIII/\SiII}
 \end{array}\right\rbrace = \text{A} b + \text{B} $}
   & $-$0.0200 & 0.0082 & 0.015 & 0.0723 & 0.2632 & 0.783\\ 
&  & 0.0057 & 0.0056 & 0.309 & $-$0.4308 & 0.1657 &  0.009\\ 
&  & $-$0.0016 & 0.0071 & 0.820 & $-$0.1708 & 0.2590 & 0.510\\ 
&  &  0.0062  & 0.0057  & 0.208 & $-$0.2200 &  0.1916 & 0.250\\ \hline
 \multirow{4}{*}{Galactic Longitude} &
 \multirow{4}{*}{$\left.\begin{array}{l}
  \text{\CIV/\CII} \\
 \text{\SiIV/\SiII} \\
 \text{\SiIV/\SiIII} \\
 \text{\SiIII/\SiII}
 \end{array}\right\rbrace = \text{A} l + \text{B} $}
 & $-$0.0008 & 0.0012 & 0.478 & $-$0.3949 & 0.2226 & 0.076\\ 
&  & $-$0.0012 & 0.0005 & 0.019 & $-$0.1754 & 0.1045 & 0.093\\ 
&  & 0.0007 & 0.0007 & 0.313 & $-$0.2916 & 0.1215 & 0.016\\ 
&  &  $-$0.0008  & 0.0008 & 0.300 &  0.0923 & 0.1174 & 0.430\\
\enddata
\tablecomments{This table contains parameters from results of linear fit models to the ionic ratios vs. latitude and longitude using parametric survival regression. The coefficients each have a standard error and $P$-values listed.}
\end{deluxetable*}

\section{Discussion} \label{section:discussion}
With the full FB sample, we can now map the kinematics and ion properties of the outflowing cool gas in the bubbles. The five new sightlines in particular provide data in the low latitude region ($b\approx15-26$\degr) close to the base of the Bubbles where the nuclear outflow is launched. Of the five new sightlines through the Fermi Bubbles, three have at least one FB HVC detection. Of the total of 15 sightlines that pass clearly through the Fermi Bubbles from the full UV FB sample 12 or 80\%\ have at least one FB HVC associated with them. In contrast, of the 54 sightlines lying \emph{outside} of the Fermi Bubbles but still in the general vicinity of the Galactic Center, only 15 or 28\%\ have HVC detections. These enhanced covering fractions support the physical association of HVCs with the Fermi Bubbles, confirming the result of \citetalias{bordoloi_2017b}. 

\subsection{Velocity Profile of the Fermi Bubbles}
\subsubsection{Velocity Profile Observations}\label{subsection:obs_vel_profile}
Our analysis of the full FB UV sample reveals that the GSR velocity of the Fermi Bubble HVCs generally decreases with increasing latitude (Figure~\ref{figure:velocity}a), with the exception of one point at $b$\s26\degr\ and $v_{\rm LSR}\sim100$ \kms. This cloud is from the UVQS J193819-432646 sightline and may represent foreground material or clouds with significant motion in the plane of the sky.

\begin{figure*}[!ht]
    \centering
    \epsscale{0.91}
    \plotone{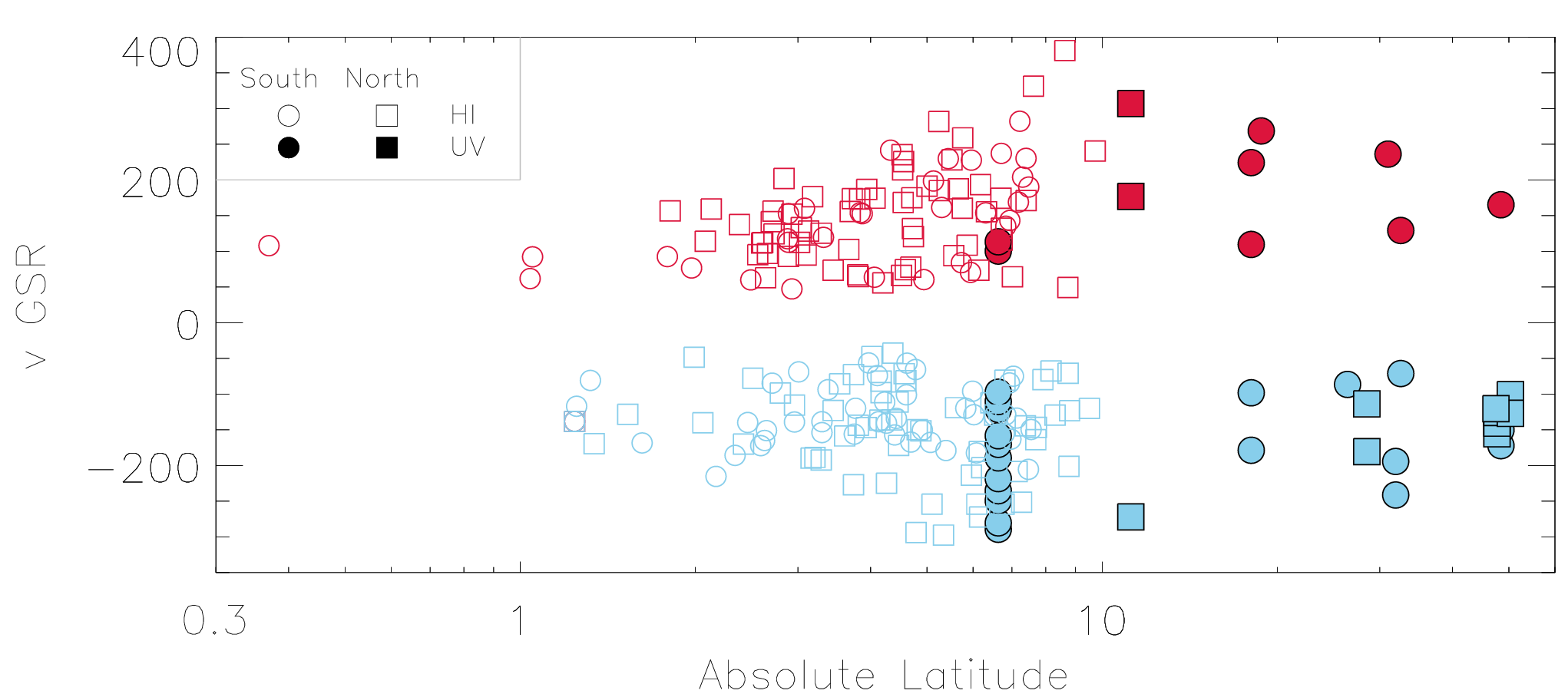}
\caption{GSR velocity vs. absolute latitude (plotted on a log scale) of the combined \HI\ and UV data from the Galactic center outflow. Red and blue symbols represent FB HVCs that are redshifted and blueshifted, respectively. Open squares and circles represent the \ion{H}{1} data from \citet{McClure_Griffiths_2013}, \citet{Di_Teodoro_2018}, \citet{Lockman_2020} from the northern and southern Fermi Bubbles, respectively. Filled squares and circles represent the UV data from the full FB sample from the northern and southern Fermi Bubbles, respectively.  \label{figure:v_UV_HI_comparison}}
\end{figure*}

Recently published \HI\ data from \citet{McClure_Griffiths_2013}, \citet{Di_Teodoro_2018}, and \citet{Lockman_2020} show clouds at low latitudes ($|b|<7$\degr) associated with nuclear outflow that are likely linked to the Fermi Bubbles. In Figure~\ref{figure:v_UV_HI_comparison} we have combined these \HI\ data with the full UV sample to show the full velocity profile of the Milky Way's nuclear wind for the first time. The figure shows GSR velocity against absolute latitude. It is worth noting that at $b\approx7$\degr, where we have overlapping coverage between the radio and UV samples, there is a remarkable agreement in the range of velocities observed, with both radio and UV samples seeing gas at up to $\pm$300\ \kms. This adds confidence to the idea that the radio and UV samples are tracing a similar population of clouds. 

Clouds near the base of the Fermi Bubbles, seen in \HI, are accelerated up to $|b|\sim7$\degr\ \citep{Lockman_2020}. After that point, the clouds (seen in the UV because the \HI\ coverage does not currently extend that high into the halo) show a change in $v_{\text{GSR}}$ of -3.3$\pm$0.1\ \kms\ per degree. As the clouds move upwards, their volume density decreases: the pathlength of the line-of-sight through the bubbles increases with increasing latitude (the bubbles are slimmer at lower latitudes) and, as seen in Figure~\ref{figure:number_density}, the number of UV detected FB HVCs per unit pathlength decreases with increasing latitude. 

We note that in Figure~\ref{figure:v_UV_HI_comparison}, the \emph{blueshifted} UV data points beyond \s30\degr\ latitude are not likely located inside of the Fermi Bubbles due to the tangential velocity constraints imposed by the geometry of the Fermi Bubbles. At a Galactic latitude of 45\degr, the near side of the Fermi Bubbles is approximately perpendicular to the line of sight, resulting in a projected velocities of zero \kms\ \citep[using the geometry outlined in][]{Miller_2016}. Nonetheless, the blueshifted FB HVCs beyond \s30\degr\ latitude may still be related to the Fermi Bubbles since they follow the trend of Fermi Bubbles clouds in $v_{GSR}$ vs. Galactic latitude. One possibility is that these FB HVCs lie just outside of the current Fermi Bubbles and are remnants of past Galactic Center outbursts  \citep[assuming that the Fermi Bubbles are a cyclic/episodic process as suggested in][]{Su_2012, Bland_Hawthorn_2019}; exploring their origin is beyond the scope of this paper.  

\subsubsection{Velocity Profile Model\label{section:model_velocity}}
The ambiguity in distances to the FB HVCs and the projection effects of their movement with respect to the plane of the sky makes it difficult to interpret the observed radial velocities. However, within the framework of some simplifying assumptions, we model the distribution of relative probabilities of FB HVC flow velocities in this section. 

We use the Fermi Bubble geometry modeled by the equations in Section 5 of \citet{bordoloi_2017b}. For details about the formulaic derivation of the geometry used in this paper, see Appendix~\ref{Appendix:Geometry}.   This geometry assumes that the Fermi Bubbles can be modeled as two cones on opposite sides of the Galactic Plane with identical opening half angles of $\alpha_{max}$. Using the results of \citet{Di_Teodoro_2018}, we assume an $\alpha_{max}$ of 70$\degr$. This choice for $\alpha_{max}$ is consistent with the boundaries of the evacuated \HI\ near the Galactic center shown on the left-side of Figure~\ref{figure:previous_work} \citep{Lockman_2016}, assuming that the lack of \HI\ is the result of the same outflow discussed in \citet{Di_Teodoro_2018}. This conic-geometry with an $\alpha_{max}$ of 70$\degr$ would \emph{not} exclude the anomalous velocity clouds beyond \s30\degr\ in latitude discussed in Section~\ref{subsection:obs_vel_profile} from the interior of the Fermi Bubbles. The exclusion of the anomalous velocity clouds is based on the \citet{Miller_2016} bubble geometry. Given the conic-geometry does not restrict the velocities of these anomalous FB HVCs and that there is limited UV data available, the anomalous velocity clouds have been included in the analysis of these models. 

The FB HVCs are assumed to be launched from the vertex of the cone uniformly over all angles to a distance, $r$, defined as the distance from the vertex to some spot on the surface of a cone: 
\begin{equation}\label{equation:r}
    r=\dfrac{\rho \text{sin} b+z}{\text{cos}\alpha}
\end{equation}
where $\rho$ is the distance from the cloud to the observer, $b$ is the Galactic latitude of the cloud, $z$ is the offset of the cone's vertex from the center of the Galaxy, and $\alpha$ is the angle of a cone on which the cloud resides (nested within the boundary of the larger cone containing the whole flow). We include the parameter, $z$, in our equations to include the possibility that the flow arises a distance away from the Galactic plane from a ring surrounding the Galactic center. However, we set $z$ = 0 in our analysis on the premise that if there was a $z$ component greater than 0, then an imprint of Galactic rotation would be expected in the flows, yet no such imprints are detected by \citet{Di_Teodoro_2018}. 

Using this geometry, we compute a differential probability $\Delta P$ per unit change in $v_{flow}$, where $v_{flow}$ is the deprojected velocity of the FB HVCs. The differential probability, $\Delta P$, is the probability of finding an individual cloud at some location within the Fermi Bubbles, which we have defined as equivalent to the density of clouds. To define this density, we consider the volume between two nested cones, one with an opening half angle $\alpha$ and another with this angle equal to $\alpha + \Delta\alpha$. The path length of the sightline that penetrates the space between these cones is given by the term $(d\rho/d\alpha) \Delta\alpha$.  The volume density of the clouds in this volume would be proportional to the inverse square of the distance, $r$, if the clouds maintain their integrity and move at constant velocity.  If the clouds are not moving at constant velocity, the density would be changed in proportion to the inverse of this velocity. Hence, the density of clouds with conserved mass should be proportional to $r^{-2}/v_{flow}$, where $v_{flow}$ is defined as
\begin{equation}
v_{flow}=\dfrac{v_{GSR}}{\text{cos} \beta},
\end{equation}
where $\beta$ is the angle between the flow direction and the observing direction. 

The differential probability is therefore defined as 
\begin{equation}\label{equation:probability}
\Delta P=\dfrac{|(d\rho/d\alpha)(d\alpha/d v_{flow})|r^{q}}{v_{flow}}\Delta \alpha,
\end{equation}
where $q$ is a radial index that quantifies how much of the clouds' mass is conserved as they travel to larger $r$. A value of $q$=\n2 describes clouds that have their mass conserved as they flow to higher $r$ and a value of q\textless\n2 describes clouds that are being eroded by the flow. For each value of $\alpha$, one can evaluate the transformation from $v_{flow}$ to $v_{obs}$ from the dot product of the unit vectors that represent the flow direction and the sightline direction. 

Figure~\ref{figure:v_flow_probabilities} shows the sums of the probabilities, $\Delta P$, as a function of distance from the Galactic center and flow velocity. The panels on the left represent the \HI\ data while the panels on the right represent the UV data. Clouds that conserve mass as they move through the Fermi Bubbles are represented in the top panels ($q=-2$) and clouds that lose mass as they move through the Fermi Bubbles are represented in the bottom panels ($q=-4$). There are significantly fewer UV FB HVC detections when compared to the number of detected \HI\ Galactic center clouds, giving the UV panels a less continuous appearance than the \HI\ panels.   

We conclude from the \HI\ data that the most likely flow velocities for values of $q=-2$ and $q=-4$ span a range of 60-250 \kms\ and 60-300 \kms, respectively, for $r$\textless1 kpc, and mostly the upper portion of this range for 1\textless $r$\textless2 kpc (\frenchspacing{i.e. } red, orange, and yellow portions of the figures). The clouds detected in the UV at 1\textless $r$\textless2 kpc have most probable velocity ranges that generally agree with those seen in the \HI\ data: 100-200 \kms\ for $q=-2$ and 125\s400 \kms\ for $q=-4$. The most probable velocity ranges in the UV data either accelerate or coast to 300-425 \kms\ for 2\textless $r$\textless 6.5 kpc in both q values. 

We note that the high probabilities at \s6.5 kpc are driven mainly by three sightlines: RBS 1768, MS-ZNG1, and Mrk 1392. Each of these sightlines has an absolute Galactic latitude of 48-50\degr\ and \emph{all} of their detected FB HVCs are blueshifted. It is therefore worth noting that in a more realistic geometry like that of \citet{Miller_2016} these clouds at \s6.5 kpc would be moving too fast to be within the gamma-ray defined Fermi Bubbles, as discussed in Section~\ref{subsection:obs_vel_profile}. Since these FB HVCs are driving the trend of acceleration and are moving too fast to be confined within the Fermi Bubbles, it is likely that $v_{flow}$  shown in Figure~\ref{figure:v_flow_probabilities} at 6.5 kpc has a range lower than 300-425 \kms. We therefore conclude that the UV data likely favor a coasting phase within the Fermi Bubbles between 2 and 6.5 kpc. 

\begin{figure*}[!ht]
    \centering
    \epsscale{1.2}
    \plotone{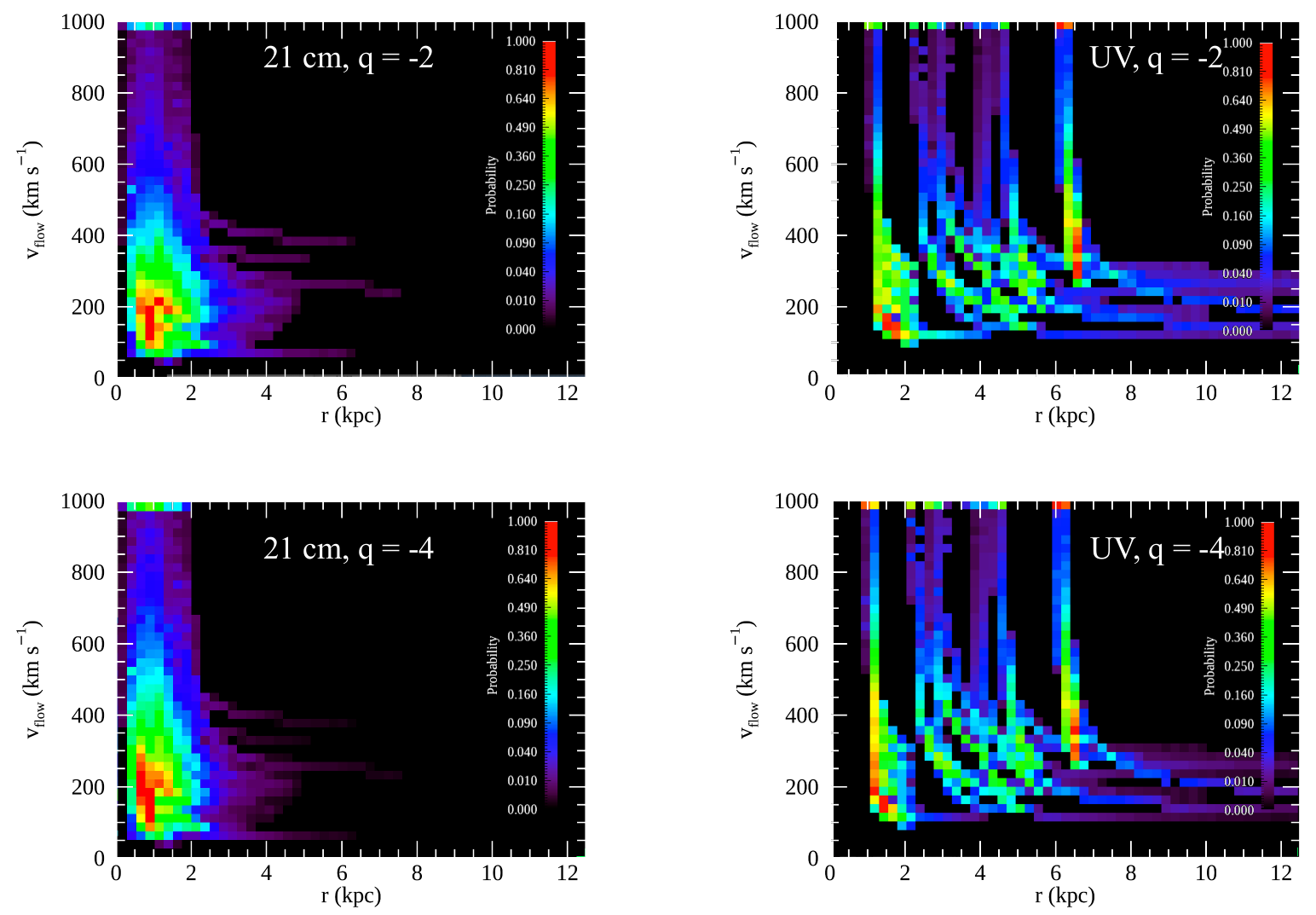}
\caption{Results from biconical modeling of the outflow kinematics. The plots show the probability of the outflow having a given combination of r, $v_{flow}$ given the H I data (left) and the UV data (right). Two $q$ parameters are used, $q=-2$ (top) and $q=-4$ (bottom) for clouds maintaining mass and clouds losing mass as they travel through the Fermi Bubbles, respectively.   \label{figure:v_flow_probabilities}}
\end{figure*}

\subsection{Gas Phases in the Fermi Bubbles}

Our UV absorption-line data constrain the thermal phase of the cool gas in the Fermi Bubbles. In principle, the gas could be in a multi-phase state where the high and low ions reside in distinct volumes at different densities and temperatures, or in a single-phase state where the high and low ions are co-spatial at a single density and temperature. Single-phase states could arise in several scenarios, including photoionization equilibrium, collisional ionization equilibrium, non-equilibrium photoionization, and non-equilibrium collisional ionization. In non-equilibrium collisional ionization, high ionization states can be observed at lower temperatures than would hold under equilibrium conditions when the cooling timescale is shorter than the recombination timescale, leaving frozen-in ionization \citep{Kafatos_1973}.

Although the ionizing environment of the Fermi Bubbles could be very different than the rest of the Milky Way halo, a useful comparison sample is provided by other Galactic HVCs (away from the Galactic Center). Analysis of UV observations including the high ion \OVI, has led several groups to favor multi-phase models for Galactic HVCs, in which the high ions arise in conductive interfaces or turbulent mixing layers around the cool gas cores \citep{Sembach_2003, Fox_2004, Fox_2005, Ganguly_2005, Collins_2005, Bordoloi_2017a}. In such a multi-phase scenario, little or no velocity centroid offset may exist between the low ions and high ions, even though the two reside in different physical regions, but a difference in $b$-values would be expected since their thermal broadening components would be different \citep{Bohringer_1987, Borkowski_1990}.

In the Fermi Bubble HVCs, the high-ion to low-ion velocity offsets allow us to test this multi-phase scenario. In Table~\ref{table:subtraction} we compared the average centroid velocity difference $\langle\Delta v\rangle$ between different ions (high, mid, and low).  The high-low ion pairs have an offset average of 9.3 \kms\ with a standard error of the mean of 1.2 \kms. The low-low ion pairs are a good comparison sample to the high-low ion pairs since we expect low ions to live with other low ions and there is a statistically significant sample of low ions. The low-low ion pairs have an offset average of 4.0 \kms\ with a standard error of the mean of 0.5 \kms. Even though the low-low ion pairs have an average offset that is half of that of the high-low ion pairs, we note that the standard deviations of their respective $\langle\Delta v\rangle$ are 5.5~\kms\ and 9.4~\kms. Such large standard deviations suggest that the high-low ion pairs may have similar average offsets as the low-low ion pairs. This suggests that the Fermi Bubble HVCs likely have a single-phase structure, conductive interfaces, or turbulent mixing layers.

Previous work with high resolution STIS data indicates \citep[e.g.][]{Keeney_2006, Zech_2008, Savage_2017} that a portion of the FB HVCs have a multiphase structure. The $b$-value distribution plot of the full Fermi Bubble sample in Figure~\ref{figure:b_values} does not obviously distinguish between multiphase and single-phase FB HVCs, making such plots difficult to interpret. Individual analysis of the FB HVCs, preferably with FUSE data covering the \OVI\ absorption and high resolution STIS data, would be necessary to determine the phase state of the individual clouds in the full Fermi Bubble sample; a task that is beyond the scope of this survey paper.




\subsection{Cooling and Recombination Timescales}
The lack of trends in the ion ratios against latitude shown in Figures~\ref{figure:ion_ratios}a-d and discussed in Section~\ref{section:ion_ratios} could indicate that
the clouds do not spend enough time in the Fermi Bubbles to cool, as suggested by \citetalias{Karim_2018}. 

\citetalias{bordoloi_2017b} show that it takes FB HVCs \s6--9 Myr to travel from the center of the galaxy to the top of the Fermi Bubbles. We can calculate the cooling time using:
 \begin{equation}\label{eq:tcool}
 t_{\rm cool} = 3kT/n\Lambda(T,Z),
 \end{equation}
 where $k$ is the Boltzmann constant, $T$ is the initial temperature of the gas in Kelvin, $n$ is the initial density of the gas in cm$^{-3}$, and $\Lambda(T,Z)$ is the temperature and metallicity dependent cooling function in erg cm$^{3}$ s$^{-1}$ \citep{Heckman_2002, Bordoloi_2017a}. The clouds probed by the UV absorption lines under study have temperatures of \s$10^{4}-10^{5}$ K. We estimate the density of the gas using the only Fermi Bubble HVC with a detailed \emph{Cloudy} photoionization model, the FB HVC at $v$=$-$172 \kms\ toward the quasar 1H1613-097.  This FB HVC has a metallicity of [O/H]$\gtrsim-0.54$ \citepalias{bordoloi_2017b}, an ionization parameter of log\,$U\ge-2.9$,
 and a derived gas density of log\,($n_{\rm H}$[cm$^{-3}$])=$-$1.2 or 
 $n_{\rm H}=0.063$\,cm$^{-3}$. The cooling function is derived for collisional ionization equilibrium (CIE) in \citet{Sutherland_1993} (see their Figure 8 and Table 6). We have calculated $t_{\rm cool}$ for four cases: temperatures of $T$=10$^4$ and $T$=10$^5$ K, and metallicities of [Fe/H]=$-$0.5 and 0.0 (solar). We chose a metallicity of [Fe/H]=$-$0.5 because it is the closest model in \citet{Sutherland_1993} to the \citetalias{bordoloi_2017b} FB HVC, and we chose the solar metallicity case since it is a reasonable expectation for gas in the metal-rich central region of the Milky Way. From these calculations, we obtain cooling timescales of 0.02-0.35 Myr, much shorter than the 6--9 Myr travel time of a gas cloud in the bubbles. 


We calculate the recombination timescale using
 \begin{equation}\label{eq:trec}
 t_{\rm rec} = (\alpha_{\rm rec} n)^{-1},
 \end{equation}
where $\alpha_{\rm rec}$ is the recombination coefficient. If $t_{\rm cool}\textless\textless t_{\rm rec}$, then the FB HVCs may have a frozen in ionization level. We obtain temperature dependent values of $\alpha_{\rm rec}$ for transitions from \CIV\ to \CIII\ ($\alpha_{\rm rec}$ of $1.53$, $3.20\times10^{-11}$ for $10^{4}$ K and $10^{5}$ K, respectively) and \CIII\ to \CII\ ($\alpha_{\rm rec}$ of $0.602$, $2.91\times10^{-11}$ for $10^{4}$ K and $10^{5}$ K, respectively) from the NORAD Atomic Data site\footnote{Nahar OSU Radiative Atomic Data website https://norad.astronomy.osu.edu}. We also use the same density used to calculate $t_{\rm cool}$.  From these calculations, we obtain $t_{\rm rec}$ values of 0.117 and 0.033 Myr for $10^{4}$ K and $10^{5}$ K, respectively. The calculated recombination timescales are broadly consistent with the cooling timescales but much shorter than the gas travel time through the Bubbles. This suggests the high ions are not experiencing frozen-in ionization, and instead may be created \textit{in situ} in the flow, either by photoionization or collisional ionization. Detailed ionization modeling would be worthwhile to investigate their origin.


\section{Conclusions} \label{section:conclusions}
Using new UV spectra from \emph{HST}/COS, we have studied five quasar sightlines passing at low latitude near the Galactic Center. Four of the five pass through the southern Fermi Bubble and the other passes just outside. Using this dataset we have mapped the kinematics and ion properties of the Fermi Bubbles absorption components. Our results can be summarized as follows.

\begin{enumerate}

\item FB HVCs are detected in three of the four sightlines that pass through the Fermi Bubbles, whereas none are detected in the sightline passing outside. Combining our data with previous quasar sightlines through the Fermi Bubbles, we find that 12 out of 15 or 80\% of Bubble sightlines have at least one FB HVC detected. This high covering fraction supports the interpretation that the FB HVCs trace gas physically associated with the bubbles.

\item We do not observe any asymmetries in the maximum absolute velocity of clouds between the northern and southern Fermi Bubbles. The dependence of velocity on Galactic latitude is similar in both hemispheres.

\item We compute the velocity profile of the outflowing cool gas by comparing $v_{\rm GSR}$ with absolute latitude for the UV-absorbing and the \HI-emitting FB HVCs. We also model the deprojected flow velocity of the FB HVCs up to 6kpc. We find that there is no evidence for deceleration of the outflow up to $|b|\approx40$\degr. 

\item We note the presence of several blueshifted anomalous-velocity FB HVCs located at Galactic latitudes of $\ge$30\degr. The geometry of the Fermi Bubbles would require the projected LSR velocities of these FB HVCs to be significantly lower than is measured since they would be moving perpendicular to the line of sight at latitudes of 45\degr. This indicates that they are likely located outside of the Fermi Bubbles. These anomalous-velocity FB HVCs could be the remnants of a past Galactic center outburst.

\item The incidence of FB HVCs (number of components per sightline) is high at $|b|<10$\degr\ and rapidly decreases until $|b|\approx$40--50\degr, where it rises again as the upper boundaries of the Fermi Bubbles are intersected. The decrease can be understood as the FB HVCs decreasing in volume density as they flow up into the halo. The incidence of FB HVCs is relatively constant with longitude. 

\item The velocity centroid offsets from high-low ion pairs and low-low ion pairs have similar distributions, as we confirm using a two-sided K-S test.  Since the high-low ion pairs have the same distribution to the low-low ion pairs (which are expected to live with each other), this result is evidence for a single-phase structure, conductive interfaces, or turbulent mixing layers.

\item The Fermi Bubble HVCs have $b$-values with similar distributions for the high (\SiIV\ and \CIV) and low ions (\SiII, \CII, and \SiIII), as we confirm using two-sided KS tests. In both cases the mean $b$-value is \s22 \kms.




\end{enumerate}

{\it Acknowledgements.}
We would like to thank David French and Elaine Frazer for their valuable help in visualizing the data. We would also like to thank John Chisholm for his valuable conversations on quasar outflows. Support for \emph{HST} programs 13448 and 15339 was provided by NASA through grants from the Space Telescope Science Institute, which is operated by the Association of Universities for Research in Astronomy, Inc., under 
NASA contract NAS 5-26555. The Green Bank Observatory is a facility of
the National Science Foundation, operated under a cooperative agreement by
Associated Universities, Inc. The GBT data were obtained under Program GBT18A-221.\\
\\
{\it Facilities:} HST (COS) and GBT\\   
\\
{\it Software:} VPFIT, IDL, R, and Python

\appendix

\section{HST/COS Absorption Line Spectra of Five New Sightlines}\label{appendix:big_table}

In Figure~\ref{figure:remaining_fits} we present the UV data for four of the five new quasar sightlines.  This figure is a continuation of Figure~\ref{figure:fits}. We shade each high-velocity absorption component that is associated with an FB HVC detected in pink or blue for redshifted and blueshifted components, respectively. We also include tickmarks in each panel to indicate the velocity of detected FB HVCs. We present all of the identified FB high velocity absorption components from the five new sightlines in Table~\ref{table:detected_lines}.  This table lists each absorption component's velocity centroid, $b$-value, column density, and line ID. The initial groupings of gas clouds (based on the visual alignment of their velocity centroids) are separated by horizontal lines. Each absorption feature determined to be part of the same cloud is assigned a number, as indicated in the Line ID column of Table~\ref{table:detected_lines}.  Upper-case letters in the Line ID column of Table~\ref{table:detected_lines} indicate that the absorption is not matched to any other absorbers (NM), or matched to some, but not all absorbers in a high-velocity cloud (N). It is worth noting that two of the three absorption features that are labeled N have low $\sigma$ values.

\begin{figure*}[ht!]
\centering
     \subfloat[UVQS J185649-544229]{\includegraphics[width=1.0\textwidth]{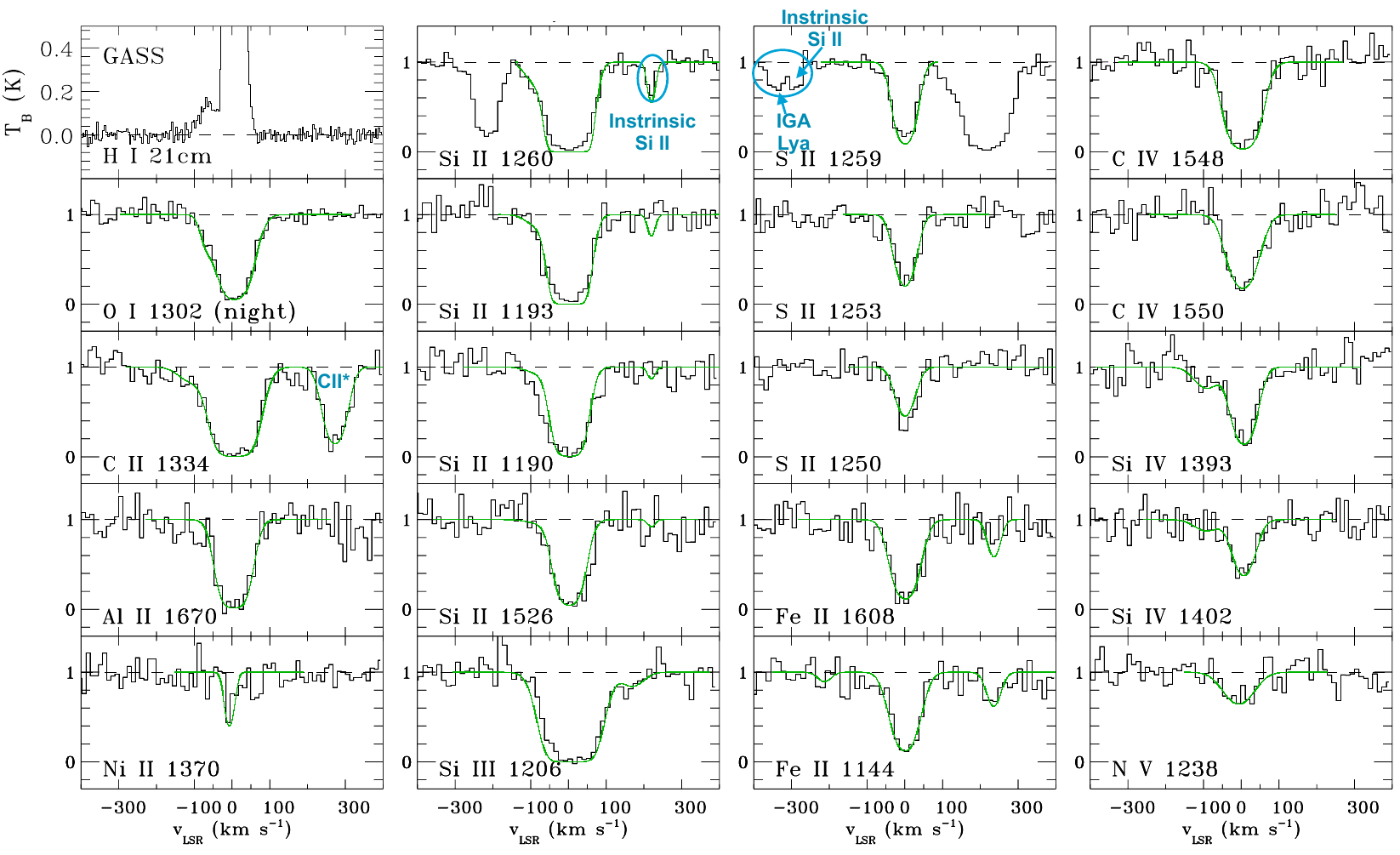}}\\
      \subfloat[UVQS J191928-295808]{\includegraphics[width=1.0\textwidth]{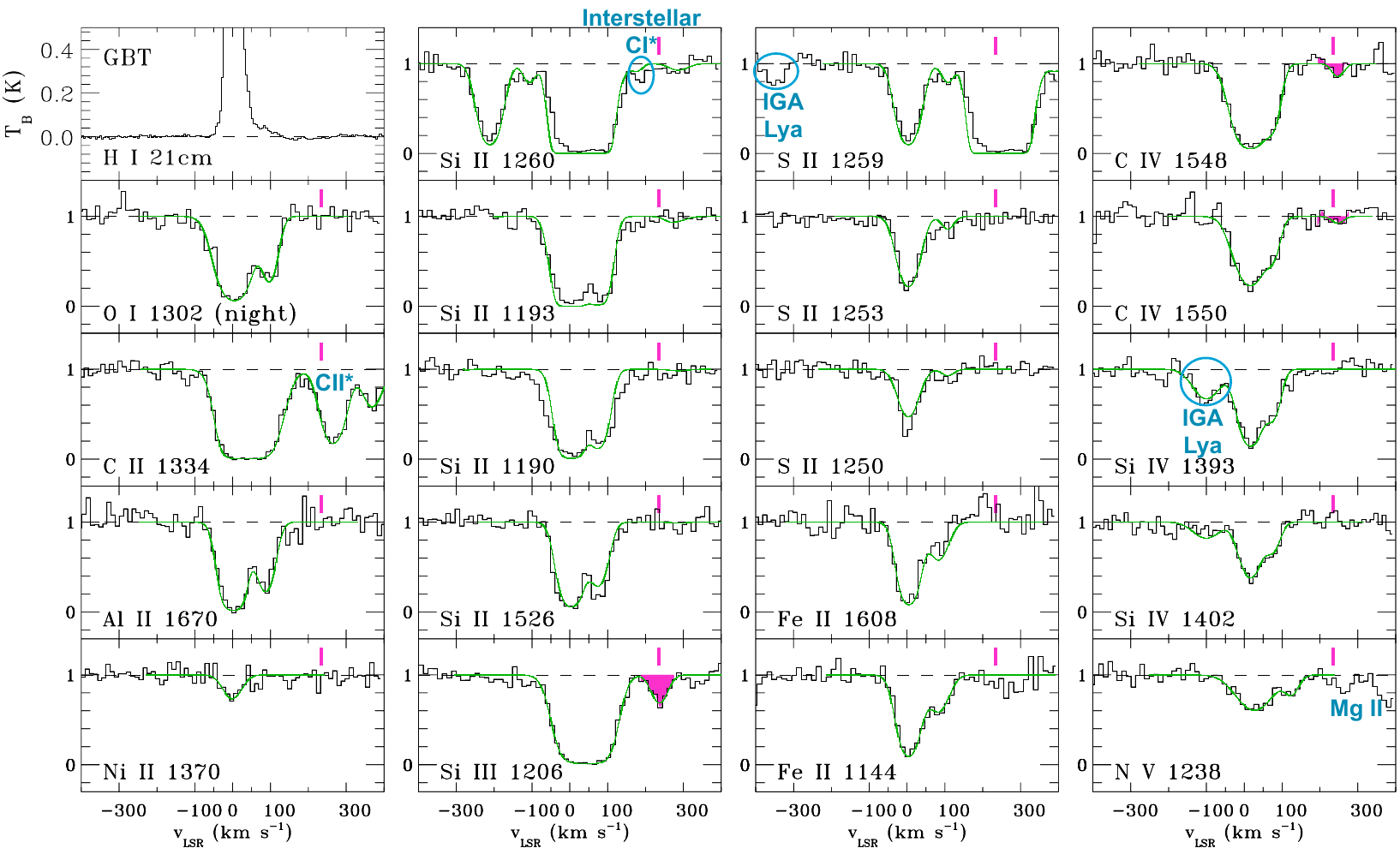}}
          \caption{Continued from Figure~\ref{figure:fits}.}
\end{figure*}
\begin{figure*}
\centering
\ContinuedFloat
     \subfloat[UVQS J192636-182553]{ \includegraphics[width=1.0\textwidth]{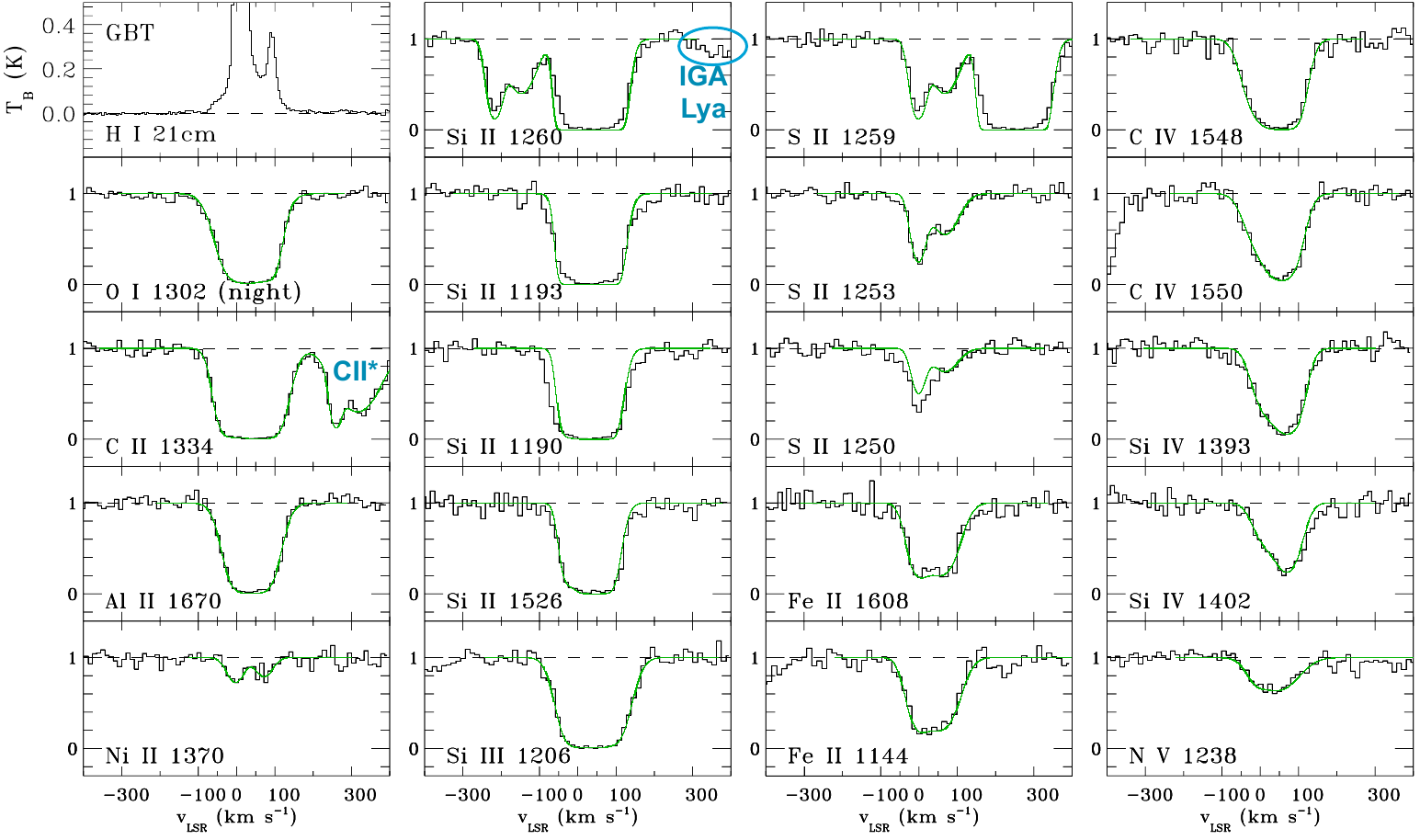}}\\
      \subfloat[UVQS J193819-432646]{\includegraphics[width=1.0\textwidth]{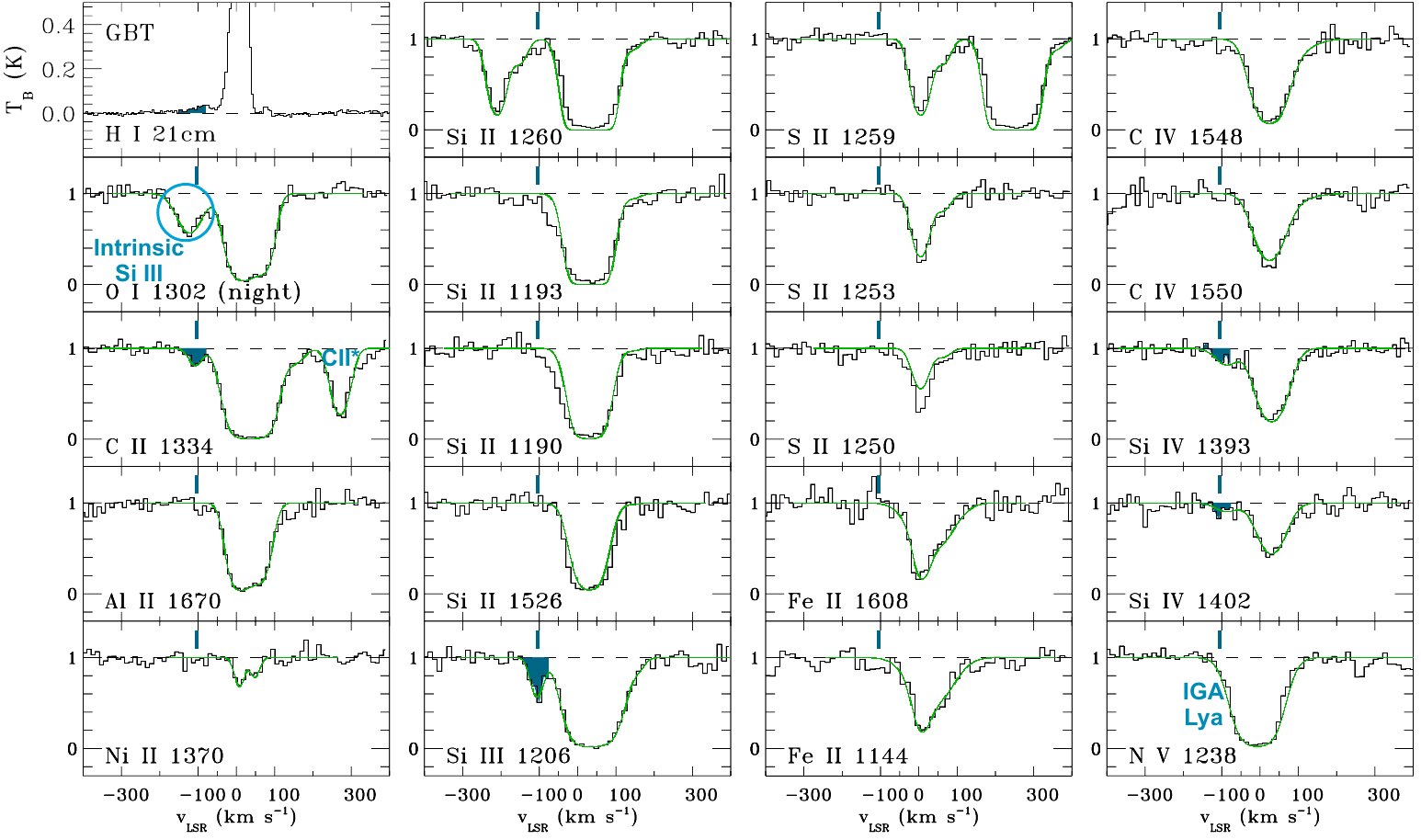}}
                \caption{Continued from Figure~\ref{figure:fits}.}\label{figure:remaining_fits}
\end{figure*}

\clearpage


\begin{center}
\begin{longtable*}{ccccccccccc}
\caption{High Velocity Absorption Line Parameters \label{table:detected_lines}}\\
\hline \hline
Sightline & Ion & $v_{\rm c\ LSR}$ & $\pm$ & error including  &  $b$-value  &  $\pm$  &  log\,$N$  & $\pm$ & $\sigma$ & Line \\
  &   &  (\kms)\tm{a} &   & systemic error\tm{b} & (\kms) &   &   &  & & ID\tm{c} \\
\hline
UVQS J185302-415839 
  & \OI &  $-$125.4 &  3.8  &  8.4  &   16.7  &  5.2  &  13.70  &  0.12   & $^{4.14}_{3.14}$  &  1 \\
  & \CII &  $-$120.2 &  2.1  &  7.8  &   33.0  &  3.5  &  14.10  &  0.04   & $^{11.33}_{10.33}$  &  1 \\
  & \AlII &  $-$122.6 &  8.5  &  11.3  &   38.2  &  12.5  &  12.41  &  0.12   & $^{4.16}_{3.16}$ &  1 \\
  & \SiII  &  $-$117.4 &  1.4  &  7.6  &   31.8  &  2.0  &  13.11  &  0.02   & $^{18.42}_{17.42}$  &  1  \\
  & \SiIII &  $-$125.7 &  1.6  &  7.7  &   23.8  &  2.0  &  12.93  &  0.05   & $^{9.79}_{8.79}$   &  1  \\
  & \CIV &  $-$122.9 &  10.0  &  12.5  &   52.7  &  11.9  &  13.66  &  0.10  &  $^{4.94}_{3.94}$  &   1  \\
  & \SiIV &  $-$118.9 &  2.0  &  7.8  &   14.5  &  2.7  &  12.67  &  0.07   &  $^{6.45}_{5.45}$  &  1  \\ 
  & \CII  &  $-$207.4 & 8.7   &  11.5  &   41.2  &  15.1  &  13.49  &  0.13   &  $^{3.81}_{2.81}$ &  2  \\
  & \SiIII &  $-$197.4 &  3.2  &  8.1  &   25.4  &  4.7  &  12.33  &  0.07   &  $^{6.96}_{5.96}$  &  2   \\ \cline{2-11}
  
  & \OI  &  106.2 &  3.0  &  8.1  &   9.5  &  4.1 &  13.44  &  0.17   & $^{3.14}_{2.14}$  &  N   \\
  & \CII &  129.2 &  2.4  &  7.9  &  30.1   &  3.0  &  14.05  &  0.04   &  $^{11.24}_{10.24}$ &  3  \\
  & \AlII &  136.9 &  14.5  &  16.3  &   67.1  &  20.2  &  12.74  &  0.11   &  $^{4.37}_{3.37}$  &  3   \\
  & \SiII &  117.9 &  2.3  &  7.8  &   38.2  &  3.2  &  13.09  &  0.03  &  $^{13.61}_{12.61}$  &  3   \\
  & \SiIII &  148.2 &  3.2  &  8.2  &   30.7  &  3.8  &  12.80  &  0.06   & $^{7.43}_{6.43}$  &  N   \\
  & \CIV &  145.2 &  6.8  &  10.1 &   32.8 &  10.4 &  13.15 &  0.11  &  $^{4.35}_{3.35}$   &  3    \\
  & \SiIV &  137.8 &  7.2  &  10.4  &   29.6  &  10.0  &  12.52  &  0.13   &  $^{3.91}_{2.91}$   &  3   \\ \cline{2-11}
  & \SiII & 250.5  &  6.9  &  10.2  &   51.6  &  10.2  &  12.51  &  0.07   & $^{6.53}_{5.53}$  &   4  \\
  & \SiIII &  245.5 &  4.2  &  8.6  &   28.3  &  6.0  &  12.29  &  0.08  &  $^{6.01}_{5.01}$  &  4   \\  \hline 
  
UVQS J185649-544229 
  &  \SiIII  &  169.0  &  15.3  &  17.0  &  39.1  &  22.8  &  12.35  &  0.21  & $^{2.58}_{1.58}$  &  NM \\
    \hline 
UVQS J191928-295808  
  &  \SiIII   &  229.0  &  2.3  &  8.0  &  28.7  &  3.8  &  12.56  &  0.05   &  $^{9.12}_{8.12}$ &  5\tm{d}  \\
  &  \SiII  &  279.3  &  11.5  &  13.7  &  34.8  &  16.3  &  12.08  &  0.18    & $^{3.01}_{2.01}$ &  N \\
  &  \CIV  &  239.7 &  6.3   &  9.8  &  22.0  &  8.7  &  12.90  &  0.15   &  $^{3.40}_{2.40}$ &  5\tm{d}  \\ \cline{2-11}
\hline 
UVQS J193819-432646  
  &   \CII  &  $-$111.1  &  3.3  &  8.2  &  24.0  &  4.7  &  13.30  &  0.07    &  $^{6.31}_{5.31}$  &  6   \\
  &  \SiIII   &  $-$109.7  &  1.8  &  7.7  &  22.0  &  2.6  &  12.62  &  0.05    & $^{10.09}_{9.09}$  &  6  \\
  &  \SiIV  &  $-$92.3  &  6.0  &  9.6 &  39.7  &  8.9  &  12.88  &  0.08   &  $^{5.78}_{4.78}$  &  6 \\ 
\hline
\end{longtable*}
\tn{}{High velocity absorption features from the five new sightlines. Absorption features are grouped by their centroid velocities.}
\tn{a}{$v_{\rm c\ LSR}$ is the absorption centroid in the LSR frame of the absorption feature.}
\tn{b}{The velocity centroid error has been added to the zeropoint error in quadrature to obtain a total velocity centroid error.}
\tn{c}{Each absorption feature is separated into clouds according methods described in Section~\ref{vel_matching}.  The label in the cloud ID column shows these separate clouds grouped together as indicated by numbers. N indicates than an absorption feature was matched by some, but not all the lines in a cloud. 
NM indicates that no matching absorber or cloud was found.} 
\tn{d}{This FB HVC has a significance of $\ge$8.12$\sigma$ in \SiIII\ and $\ge$2.4$\sigma$ in \SiII. We chose to keep this FB HVC in our analysis because \SiIII\ has a large f-value, therefore we would expect any FB HVC detection to be strongest in \SiIII.}
\end{center}

\section{Velocity Centroid Offset Calculations vs. Cloud Matching Process}\label{appendix:centroid_offset}

The cloud matching process described in detail in Section~\ref{vel_matching} assumes that the absorption features in two ions should have matching velocity centroids to within 3$\sigma$ if they are part of the same FB HVC. 
However, if the ionization mechanism of the gas produces a large velocity offset between the low and high ions, then by including only the matched ions, we could have introduced a bias towards low $\Delta v$ values contributing to calculations presented in Table~\ref{table:subtraction}.  

To address this concern we re-calculated the $\langle\Delta v \rangle$ values for each ion pair in the full FB sample for several situations: all components (matched and unmatched), all matched components on the same grating, and all components (matched and unmatched) on the same grating. The calculations of $\langle\Delta v \rangle$ on the same grating ensure that $\langle\Delta v \rangle$ is not affected by the zeropoint offset between the gratings.  These cases have been added to the calculations of $\langle\Delta v \rangle$ for matched components in Table~\ref{table:subtraction_comparison}.

In each case (matched, all components, matched same grating, and all components same grating) the $\langle\Delta v \rangle$ values stay relatively constant for the ion pair types. We note that pairs including high ions have $\langle\Delta v \rangle$ values that are slightly higher when unmatched components are included, however, these differences are small and within the errors. When only ion pairs in the same grating are accounted for, we do see a slight drop in several the $\langle\Delta v \rangle$ values for pairs including low ions, with the most significant decrease in $\langle\Delta v \rangle$ values being in the high-low ion pairs. We note that none of these changes are within a 3$\sigma$ significance. This demonstrates that our 3$\sigma$ cutoff has not biased the centroid offset results and that the zeropoint offset is also not affecting the centroid offset results. 

\begin{deluxetable}{lcccc}[!ht]
\tablecaption{Velocity Centroid Offsets \label{table:subtraction_comparison}}
\tablehead{\colhead{Ion Pair} & \colhead{$\langle\Delta v \rangle$} & \colhead{Standard Error} & \colhead{Standard} & \colhead{Number of}\\[-8pt]
\colhead{} & \colhead{} & \colhead{of the Mean} & \colhead{Deviation} &\colhead{Pairs}\\[-8pt]
& \colhead{(\kms)} & \colhead{(\kms)} & \colhead{(\kms)} &}
\startdata
\multicolumn{5}{c}{Matched Components}\\ \hline
High-Low & 9.3 & 1.2 & 9.4 & 64\\
High-Mid & 7.8 & 1.7 & 7.8 & 22\\
Mid-Low & 8.0 & 1.1 & 7.0 & 43\\
Low-Low & 4.0 & 0.5 & 5.5 & 146\\
High-High & 11.0 & 3.2 & 12.6 & 16\\\hline
\multicolumn{5}{c}{All Components}\\\hline
High-Low & 10.8 & 1.3 & 11.0 & 75\\
High-Mid & 8.5 & 1.7 & 8.2 & 23\\
Mid-Low & 9.9 & 1.5 & 10.4 & 47\\
Low-Low & 4.2 & 0.4 & 5.6 & 154\\
High-High & 13.9 & 4.1 & 17.1 & 17\\\hline
\multicolumn{5}{c}{Matched Components; Same Grating}\\ \hline
High-Low & 6.3 & 1.4 & 6.3 & 20\\
High-Mid & 8.3 & 2.4 & 8.4 & 12\\
Mid-Low & 5.6 & 1.3 & 5.7 & 18\\
Low-Low & 2.7 & 0.9 & 3.3 & 14\\
High-High & ... & ... & ... & 0\\\hline
\multicolumn{5}{c}{All Components; Same Grating}\\\hline
High-Low & 8.4 & 1.6 & 8.2 & 27\\
High-Mid & 9.4 & 2.5 & 9.0 & 13\\
Mid-Low & 8.2 & 2.1 & 9.8 & 21\\
Low-Low & 4.5 & 1.2 & 5.4 & 21\\
High-High & 33.5 & ... & ... & 1\\
\enddata
\tablecomments{This table shows the mean velocity centroid offset, its standard error, and the velocity centroid offset standard deviation for various ion pairs, for different sub-samples of absorption.}
\end{deluxetable}

\section{Derivation of the Fermi Bubble Cone Modeling Geometry}\label{Appendix:Geometry}

Interpreting the velocity profile of the Fermi Bubbles is difficult due to projection effects and distance ambiguities. Using a conic-geometry model for the Fermi Bubbles, the \HI\ Galactic center outflow data, and the UV Fermi Bubble HVC data, we have modeled the probabilistic deprojected flow velocities within the Fermi Bubbles. Below we derive the geometry used for calculating the differential probability in Equation~\ref{equation:probability}.

The shapes of the gamma ray emission lobes shown in Figure~\ref{figure:previous_work} indicate that the northern and southern Fermi Bubbles are approximately symmetric about the Galactic plane. Thus, we treat results for negative latitudes in the same manner as for positive ones. We also propose that the origin of the clouds is at or extremely close to the Galactic center. This statement is supported by the 21-cm \HI\ data of \citet{Di_Teodoro_2018} which probes the low latitude regions of the Galactic center ($3\degr\le b\le10\degr$). The FB HVCs detected in \citet{Di_Teodoro_2018} show a lack of systematic velocity differences between negative and positive Galactic longitudes and are therefore not participating in Galactic rotation. \citet{Di_Teodoro_2018} conclude that the observed FB HVCs are likely participating in an outflow originating from the Galactic center. While this is an easy assumption to implement, in the development of equations that follow, we will retain a parameter, z, which represents the offset of a virtual vertex of a cone from the plane of the Galaxy. As long as we do not violate the constraint of seeing no signature of rotation, a small value for z may be permissible for representing an outflow from a disk of active stars surrounding the Galactic center.

We now envision that clouds are launched uniformly over all solid angles inside a cone with an opening half-angle, $\alpha_{max}$. The clouds detected by \citet{Di_Teodoro_2018} indicate that $\alpha_{max}$ could be as large as 70$\degr$ (assuming that $z=0$), a finding consistent with the lack of \HI\ shown in Figure 2 of \citet{Lockman_2016}. 

\begin{figure}[!hb]
    \centering
    \epsscale{1.1}
    \plotone{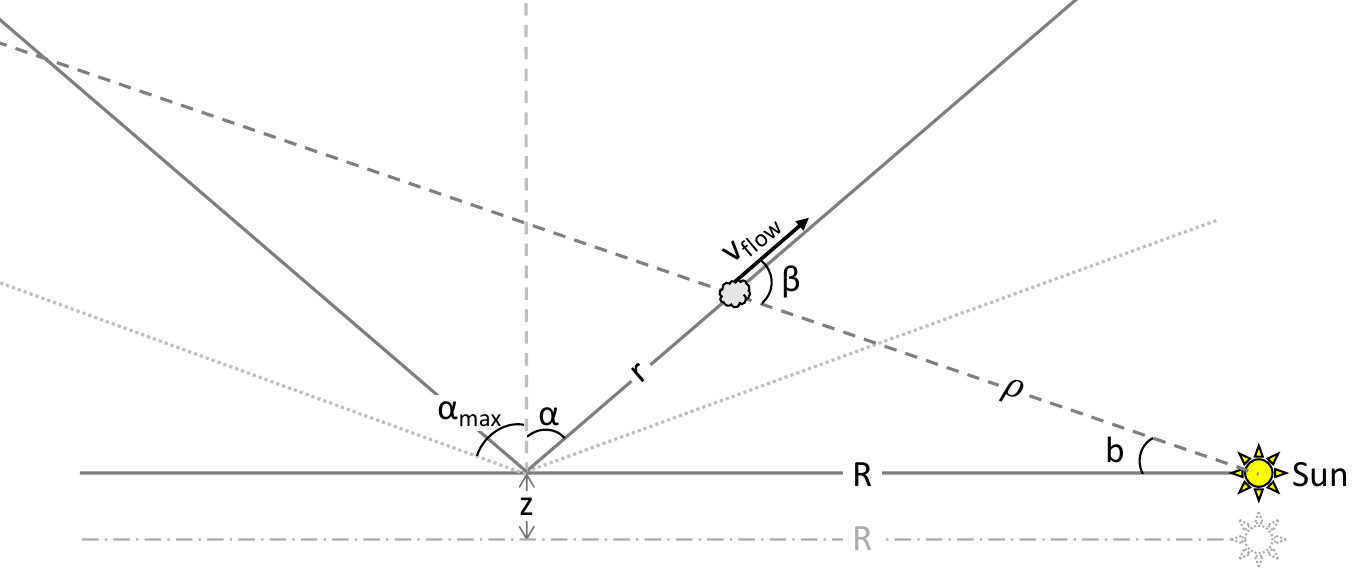}
\caption{A cartoon of the conic-geometry Fermi Bubble model. The dash-dot line is used to show the parameter z; it represents the Galactic plane if the outflow where to start a distance z off of the Galactic plane. \label{figure:Eds_model_geometry}}
\end{figure}

Drawing upon a development of equations in Sections 5.2 and 5.3 of \citet{bordoloi_2017b}\footnote{We have made some changes in notation to the equations in \citet{bordoloi_2017b}: their ${\cal L}$ is equivalent to our $\rho$ and their OA/2 is equivalent to our $\alpha$.}, we state that the distance from the vertex to a spot on the surface of any cone within the Fermi Bubbles with a half angle, $\alpha$, is given by Equation~\ref{equation:r}: \begin{equation*}
    r=\dfrac{\rho sin b+z}{cos\alpha}
\end{equation*}
In Figure~\ref{figure:Eds_model_geometry} we present the geometry and parameters used in these derivations.

In Equation~\ref{equation:r}, the distance $\rho$ of a cloud to an observer located at a distance of R = 8.3 kpc from the Galactic center is given by one of the solutions to the quadratic equation:
\begin{equation}\label{equation:quadratic}
    \text{A}\rho^{2}+\text{B}\rho+C=0
\end{equation}
where
\begin{subequations}
\begin{gather}
    \text{A}=\dfrac{\text{sin}^{2}\,b}{\text{cos}^{2}\,\alpha}-1\\
    \text{B}=\dfrac{2z\,\text{sin}\,b}{\text{cos}^{2}\,\alpha}+2R\,\text{cos}\,l\,\text{cos}\,b-2z\,\text{sin}\,b\\
    \text{C}=\left(\dfrac{1}{\text{cos}^{2}\,\alpha}-1\right)\,z^{2}-R^{2}.
\end{gather}
\end{subequations}
The two roots to Equation~\ref{equation:quadratic} express the distances to the near and far sides of the cone intercepted by a sightline with the coordinates $b$ and $l$. If $\text{B}2-4\text{AC} < 0$, $\alpha$ is so small that there is no solution because the sight line misses the cone. 

From our vantage point, a parcel of gas moving radially along the surface of a cone within the Fermi Bubbles at an outward velocity, $v_{flow}$, will be observed at a reduced velocity $v_{GSR}=v_{flow}\,\text{cos}\,\beta$, where $\beta$ is the angle between the flow direction and the observing direction. The projection of one of these directions on the other is given by
\begin{equation}
\text{cos}\,\beta=\dfrac{\rho-R\,\text{cos}\,l\,\text{cos}\,b+z\,\text{sin}\,b}{r}.
\end{equation}
For small values of r and large values of $\alpha$ on the near side of the cone, $\text{cos}\,\beta$ can revert to negative values and cause a sign reversal between $v_{GSR}$ and $v_{flow}$. For this reason, we must accept the possibility that observations could possibly represent outflowing motions ($v_{GSR}\textless 0$) or infalling high-velocity clouds that are unrelated to the outflow ($v_{GSR}\textgreater 0$). In our analysis, we treat every high velocity cloud as if it were relevant to the outflow and ignore the unavoidable prospect that some of them are contaminating our sample, since they may belong to the population of HVCs seen elsewhere at high Galactic latitudes.

Using the geometry described above, we calculate a differential probability, $\Delta P$, per unit change in $v_{flow}$ for relevant values of $v_{flow}$ and $r$ in Section~\ref{section:model_velocity}. $\Delta P$ can be calculated for any differential element $\Delta \alpha$ between the minimum $\alpha$ for a sight line to penetrate the cone and $\alpha_{max}$. 

\bibliographystyle{aasjournal}

\bibliography{5agn_sightlines}

\end{document}